\documentclass[10pt,twocolumn]{article}
\pdfoutput=1
\usepackage{amsmath}
\usepackage{amssymb}
\usepackage{url}
\usepackage{graphicx}

\usepackage{cite}

\usepackage{color} 

\usepackage[labelfont=bf,labelsep=period,justification=raggedright]{caption}

\makeatletter
\renewcommand{\@biblabel}[1]{\quad#1.}
\makeatother
\title{Identifying Overlapping and Hierarchical Thematic Structures\\in Networks of Scholarly Papers:\\A Comparison of Three Approaches}
\author{Frank Havemann$^{1,\ast}$  \and Jochen Gl{\"a}ser$^{2}$  \and Michael Heinz$^{1}$ \and Alexander Struck$^{1}$}
\begin{document}
\maketitle
 % Title must be 150 characters or less
\begin{flushleft}
 %{\Large
 %\textbf{Identifying overlapping and hierarchical thematic structures in networks of scholarly papers: A comparison of three approaches}
 %}
 % Insert Author names, affiliations and corresponding author email.
 %\\
 %Jochen Gl{\"a}ser$^{1,\ast}$, 
 %Frank Havemann$^{2}$, 
 %Michael Heinz$^{2}$,
 %Alexander Struck$^{2}$
 %\\
 %\bf
{1} Institut f{\"u}r Bibliotheks- und Informationswissenschaft, Humboldt-Universit{\"a}t zu Berlin, Berlin, Germany 
\\ 
 %\bf
{2} Zentrum Technik und Gesellschaft, Technische Universit{\"a}t Berlin, Berlin, Germany
\\ 

$\ast$ E-mail: Frank (dot) Havemann (at) ibi.hu-berlin.de
\end{flushleft}

% Please keep the abstract between 250 and 300 words
\section*{Abstract}
We implemented three recently proposed approaches to the identification of over­lapping and hierarchical substructures in graphs and applied the corresponding algorithms to a network of 492 information-science papers coupled via their cited sources. 
The thematic substructures obtained and overlaps produced by the three hierarchical cluster algorithms were compared to a content-based categorisation, which we based on the interpretation of titles and keywords. We defined sets of papers dealing with three topics located on different levels of aggregation: \textit{h-index}, \textit{webometrics}, and \textit{bibliometrics}. We identified these topics
with branches in the dendrograms produced by the three cluster algorithms and compared the overlapping topics they detected with one another and with the three predefined paper sets. We discuss the advantages and drawbacks of applying the three approaches to paper networks in research fields.

\section{Introduction}

The delineation of scientific fields is a pertinent problem of science studies in general and biblio­metrics in particular (cf. e.g. van Raan, 2004 \cite[p. 39]{vanRaan2004measuring}). Bibliometric research has shown that clusters in networks of papers do not have natural’ boundaries (cf. Zitt \textit{et al.}, 2005 \cite{Zitt2005relativity}). This is why fields must be delineated by applying thresholds for parameters, which are chosen arbitrarily in terms of `good structures' for the pur­poses of the analysis at hand (cf. e.g. references \cite{janssens2008hmi,klavans2011using}).

However, the problem of delineation might be a consequence of the overlap of thematic structures. The overlap of themes in publications is well known to science studies. Sullivan \textit{et al.} (1977) \cite[p. 235]{sullivan1977cocitation} observed that in the literature of the field of weak interaction half of the references were articles outside the specialty. Amsterdamska and Leydesdorff (1989) \cite[p. 461]{amsterdamska1989citations} provide an example of an article that targeted two different specialties at once. If disjoint clusters of co-cited sources (Marshakova 1973 \cite{marshakova1973ssm}, Small 1973 \cite{Small1973cocitation}) are projected forward to their citing papers, the clusters of citing papers inevitably overlap---a phenomenon that has never been explored by bibliometrics. Taken together, these observations suggest that the sciences consist of numerous fields of different sizes that partially or totally overlap, i.e. feature hierarchies as well as mutually overlapping `neighbours' with fuzzy boundaries.

If thematic structures have boundaries that are hidden by their overlaps, delineation is not impossible in principle but rather depends on tools that enable the identification of overlapping fields and topics.

So far, only one such tool, namely co-citation analysis, has been applied to the delineation task. However, it assumes disjoint source clusters and locates thematic overlaps only in citing papers. 
This unrealistic assumption makes it unsuitable to detect overlapping topics. An attempt to obtain overlapping thematic structures by singular value decomposition (SVD) of paper-source matrices failed because SVD produces---at least if arbitrary thresholds are avoided---as many thematic substructures as there are papers \cite{Mitesser2008mdr}, which again is an unrealistic assumption.

The aim of this paper is to introduce and preliminarily assess three algorithms
for the identification of overlapping thematic structures in networks of papers. We derived these algorithms from three recently proposed approaches to the detection of overlapping and hierarchical substructures in networks---which in network analysis are called \textit{communities}. For a concise description of the current state of finding communities in networks see the introduction of reference \cite{Lancichinetti2011finding}. Our selection and specification of the general approaches is based on the assumption that the thematic substructures both overlap and build hierarchies.

We further had to take into account the information utilised by the different approaches. Thematic structures can be determined top-down using global information or bottom-up using either global and local or only local information. This corresponds to different ways in which scientific perspectives are used in the construction of thematic structures. Since the production of contributions to scientific knowledge is based on the interpretation of that knowledge by individual producers \cite{glaser2006wissenschaftliche}, thematic structures in paper sets are always constructed from the individual perspectives of the authors. A bottom-up approach using only local information enables the reconstruction of thematic structures from the perspective of those contributing knowledge to these themes. The use of global information in the top-down or bottom-up construction of thematic structures, e.g. by spectral and modularity-based methods \cite[p. 41, p. 27]{fortunato2010community}, is akin to including the perspective of `outsiders', i.e. of authors/papers not contributing to the specific topic. Such a `democratic' procedure can be justified as well but is likely to lead to different results (for an attempt to justify the global perspective see Klavans and Boyack, 2011 \cite{klavans2011using}).

These considerations made us select three approaches that enable the identification of overlapping and hierarchical structures in networks on the basis of local information. A first approach starts from hard clusters obtained by any clustering method and fractionally assigns the nodes at the borders between clusters to these clusters (cf. e.g. Wang \textit{et al.}, 2009 \cite{wang2009adjusting}). Another approach is based on a hard clustering of links bet­ween nodes into disjoint modules, which makes no­des members of all modules (or communities) that their links belong to (cf. e.g. Ahn \textit{et al.}, 2010 \cite{ahn2010link}). The third approach constructs \textit{natural communities} of all nodes, which can overlap with each other, by applying a greedy algorithm that maximises local fitness (cf. e.g. Lanci­chinetti \textit{et al.}, 2009 \cite{lancichinetti2009detecting}).

We introduce our implemen­tations of the three approa­ches and discuss their basic features using a small benchmark graph (the karate club, Zachary, 1977 \cite{zachary1977information}) as an example. The comparative analysis applies the algorithms to a network of 492 bibliographically coupled papers published 2008 in six information-science journals. The use of information-science papers enabled the construction of paper sets of selected topics 
by manually assigning papers to the topics \textit{h-index}, \textit{webo­metrics}, and \textit{bibliometrics} on the basis of titles, abstracts, and keywords. The clustering solutions and the overlap of modules were then assessed by comparing them to the paper sets. On the basis of this comparison we discuss advantages and disadvantages of the three algorithms.

\section{Communities in Networks}

\label{Communities} In network analysis, communities are understood as cohesive subgroups of nodes separated from the rest of the graph. Thus, communities can only be found in networks if there are groups of densely interconnected nodes that are only loosely connected to each other. Most community definitions are based on these two aspects, i.e. cohesion and separation \cite[pp. 83--87]{fortunato2010community}. In order to apply algorithms for the detection of communities, cohesion and separation must be defined \cite{Tibely2011Criterions}. Owing to the continuous nature of the two properties, communities cannot be detected unequivocally. Instead, structures of varying `communityness' can be identified \cite{friggeri2011ego}.

In the case of thematic structures in networks of papers, the communities to be detected do not only overlap each other but are also hierarchically ordered. For hierarchies of communities, both cohesion and separation can be measured directly in the dendrogram. The simplest measure of separation of a community is the similarity level $s_u$ at which its branch in the dendrogram unites with another branch. The simplest measure of cohesion of a community is the similarity level $s_d$ at which its branch in the dendrogram decays into two branches but here low level means high cohesion. Thus, good communities are those with high level $s_u$ and low level $s_d$. This corresponds nicely to the usual selection of long branches as important ones. They have large differences $s_u - s_d$, i.e. are stable over relatively large similarity intervals. Using this difference, we can order branches with respect to their quality as communities, i.e. combined cohesion and separation.

In our experiments, the stability is negatively correlated with community size. Many small branches are very stable and many larger branches are very unstable. In order to find `interesting' communities, we plot branch length $s_u - s_d$ over community size and identify communities that are unusually stable for their size, i.e. are represented by branches far from the axes of the plot.

An alternative approach to community delineation associates cohesion with high internal and separation with low external degrees of community members. The internal degree $\kappa_{\mathrm{in}}(C, V_i)$ of a node $V_i$ is defined as the sum of weights of edges linking this node with nodes in community $C$, its external degree $\kappa_{\mathrm{out}} = \kappa - \kappa_{\mathrm{in}}$, where $\kappa$ is the node's total degree. Radicchi \textit{et al.} (2004) \cite{Radicchi2004defining} define a \textit{community in the strong sense} as a set of nodes all of which have higher internal than external degree. For a \textit{community in the weak sense} they only demand that the sum of internal degrees exceeds the sum of external degrees. These sums are usually referred to as the internal and external degrees of community $C$:
\begin{equation}
k_{\mathrm{in, out}}(C) = \sum_{V_i \in C} \kappa_{\mathrm{in, out}}(C, V_i).           \end{equation} 
These internal and external degrees of a community can be used to define its fitness (see the approaches `natural communities' and `fuzzification' for applications). By combining cohesion and separation, the fitness measure evaluates the quality of a community in a similar way as $s_u - s_d$ does based on a community's branch in a dendrogram.

When applied to overlapping communities, the measures used in the delineation of weak communities must take the nature of overlaps into account. Following Steve Gregory (2011) \cite{gregory2011fuzzy}, we distinguish between crisp and fuzzy overlapping communities. If a network has crisp overlapping communities, nodes either belong or don't belong to a community. Communities are fuzzy if individuals' grades of membership vary. This type of structure is appropriate for the relationship between papers and topics because most papers cover several topics in varying intensities, which led us to the application of fuzzy set theory.

Fuzzy set theory operates with membership grades that are real numbers between zero and one but does not assume that a node's grades of membership in different sets sum up to unity. A node could also be a full member in more than one community.

To determine whether a fuzzy community $C$ is a community in the weak sense we have to redefine its internal and external degree $k_{\mathrm{in, out}}(C)$ by weighting the degrees with node membership grades. With $\mu_i(C) = 1$ if $V_i \in C$ and $\mu_i(C) = 0$ otherwise, we can rewrite the definitions given above for crisp communities as 
\begin{equation} \label{fuzzy-k_in} k_{\mathrm{in}}(C) = \sum_{i, j=1}^n \mu_i(C) a_{ij} \mu_j(C) 
\end{equation} and \begin{equation} \label{fuzzy-k_out} k_{\mathrm{out}}(C) = \sum_{i, j=1}^n \mu_i(C) a_{ij} [1 - \mu_j(C)], \end{equation} 
where $a_{ij}$ is the weight of edge $(i, j)$ and $n$ the graph size. These formulae can also be used for a fuzzy community $C$ if $\mu_i(C)$ is identified with node's $V_i$ membership grade in $C$. Then $1 - \mu_i(C)$ is its membership grade in $C$'s fuzzy complement. Fuzzy set $C$ is a community in the weak sense if $k_{\mathrm{in}}(C) > k_{\mathrm{out}}(C)$.

\section{Three Approaches} 

We introduce the three approaches and explain their basic mechanisms with a simple example, namely the social network of 34 members of a karate club analysed by Zachary (1977) \cite{zachary1977information}.\footnote{using the unweighted graph: \url{http://networkx.lanl.gov/examples/graph/karate_club.html}} Members of the karate club were asked about friendship ties. The network turned out to have two central actors who, after the split of the original club, founded separate new clubs. Authors who implemented algorithms based on the three approaches applied them to the network described by Zachary. 

\subsection{Natural Communities} 

\begin{figure}[!b] 
\begin{center} 
\includegraphics[width=3in]{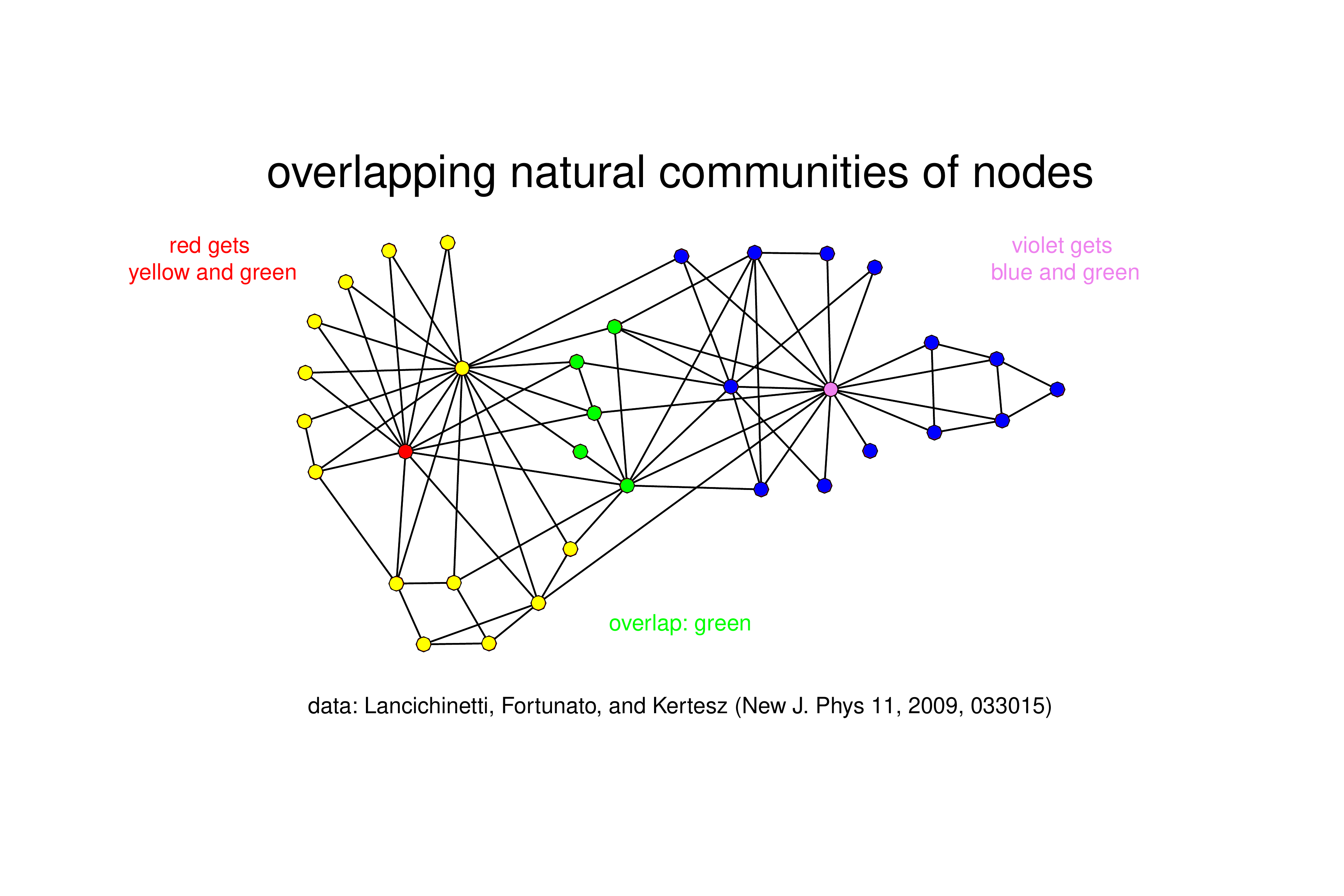} 
\end{center} 
\caption{ {\bf Natural communities.} Karate club graph with overlapping communities of two nodes~(red and violet). } 
\label{Fig-karate-LFK} 
\end{figure}

A natural community of a node can be constructed by any `greedy' algorithm which evaluates the inclusion of neighbouring nodes into the community using an appropriate metric or fitness function. If a community with a neighbour node is fitter than without it, the neighbour will be included. The essence of this local approach is that independently constructed natural communities of nodes can overlap. Figure \ref{Fig-karate-LFK} shows two overlapping communities of karate club members. On the left-hand side, the red node's community has all yellow and green nodes as its members. On the right-hand side, the violet node's community has all blue and green nodes as members. Thus, we have five (green) nodes in the overlap of both natural communities.

The idea to identify overlapping communities as sub-graphs which are locally optimal with respect to some given metric was first published by Baumes \textit{et al.} (2005)~\cite{baumes2005finding}. It can be implemented in several ways.  In the same year, Baumes \textit{et al.} \cite{baumes2005efficient} tested a combination of two greedy algorithms which both use the same metric. The \emph{initialisation} produces disjoint seed clusters, the metric of which is then improved by an iterative procedure leading to overlapping communities.

Lancichinetti \textit{et al.} (2009) \cite{lancichinetti2009detecting} combined the concept of locally optimal sub-graphs with the idea of variable resolution to enable their algorithm to reveal hierarchical community structures. They introduced a resolution parameter into their fitness function. Higher resolution results in smaller, lower in larger natural communities. The fitness function includes only local information. It is defined as the ratio of the sum of internal degrees $k_{\mathrm{in}}(C)$ to the sum of all degrees $k(C)= k_\mathrm{in}(C) + k_\mathrm{out} (C)$ of nodes in a community $C$. The denominator is taken to the power of $\alpha$, the resolution parameter:

\begin{equation} 
\label{def.fitness}
f(C, \alpha) = \frac{k_\mathrm{in}(C) }{k(C)^\alpha}.
\end{equation}  

Figure 1 displays a cover of the karate-club network obtained by Lancichinetti \textit{et al.} with a stochastic version of their algorithm for the resolution interval $0.76 < \alpha < 0.84$. Their LFM (\textbf{l}ocal \textbf{f}itness \textbf{m}aximisation) algorithm has to be repeated for all resolution levels of interest.

The construction of a scientific paper's natural community in a similarity network of papers can be interpreted as the construction of its thematic environment from its own `scientific perspective'. This idea is attractive from a conceptual point of view because it mimics the way in which scientists apply their individual perspectives when construc­ting their fields. This is why locality is a realistic assumption for topic extraction in paper networks.

At  the same time, the strictly local approach enables the local exploration of networks which are too big for global analysis like the Web or the complete citation network of scientific papers. A node's natural community is a local structure that can be constructed without knowing the whole graph. The idea to find local community structures without knowing the whole graph by using a greedy local cluster algorithm goes back to Clauset (2005) \cite{clauset2005flc}. His procedure can also be used to construct overlapping graph modules \cite{lee2011seeding}. In contrast to the resolution-depending fitness function of Lancichinetti \textit{et al.} (2009) \cite{lancichinetti2009detecting} Clauset evaluated modules with a function that does not depend on resolution.

\subsection{Link Clustering}

If links instead of nodes are clustered, nodes with more than one link can be fractional members of clusters, as figure 2 shows for the karate club. For example, vertex 1 (violet point) has four edges belonging to one and twelve edges belonging to another hard cluster of links. Thus, it has membership grades 4/16 and 12/16, respectively, in the two clusters. 

\begin{figure}[!b] 
\begin{center} 
\includegraphics[width=3in]{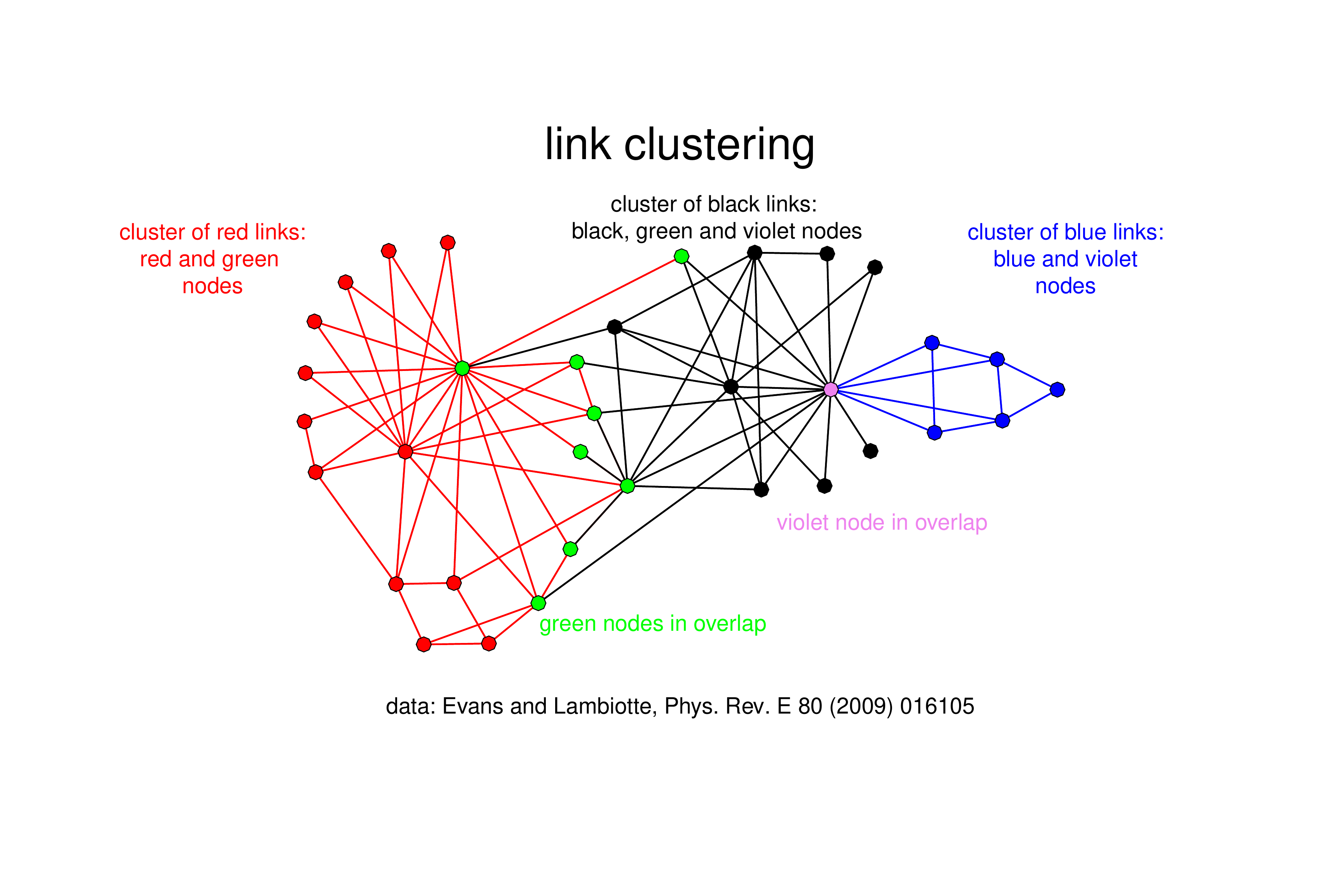} 
\end{center} 
\caption{ {\bf Hard clusters of links.} Karate club graph with overlapping node communities induced by three hard link clusters. } 
\label{Fig-karate-HLC} 
\end{figure} 

For clustering links we need a measure of link similarity. Ahn \textit{et al.} (2010 \cite[eq.~2, p.~5]{ahn2010link}) chose the Jaccard index of neighbourhoods of nodes attached to two links (a node itself is included into its neighbourhood).

In a different approach to link clustering, Evans and Lambiotte (2009) \cite{evans2009line} used the line graph of an undirected graph. To get a graph's line graph first a bipartite graph of the graph's nodes and edges is constructed by putting an edge node on each edge. The bipartite graph can then be projected onto the line graph, a graph where nodes and edges have interchanged their roles. 

Recently Ball \textit{et al.} (2011) \cite{Ball2011efficient} successfully tested an algorithm which finds overlapping node communities with a generative stochastic model of hard link clusters. Kim and Jeong (2011) \cite{kim2011map} applied the fast \emph{Infomap} \cite{rosvall2008maps} algorithm to link clustering.

The clustering of citation links instead of papers is of high interest to bibliometrics because a citation is probably the conceptually most homo­genous bibliometric unit. Since many references are referred to only once in a paper, it can be assumed that these links between the citing and the cited publication can be assigned to one theme. Even though there are many cases in which a paper cites a source for several different reasons, a citation link can be assumed to have a higher thematic homogeneity than a publication. Based on this assumption of homogeneity, citation links can be hard-clustered, which leads to overlapping clusters of papers. The membership grade of a pa­per to a module corresponds to the part of outgoing citation links of this paper within this link cluster.

We applied the \textbf{h}ierachical \textbf{l}ink \textbf{c}lustering (HLC) method suggested by Ahn \textit{et al.} (2010) \cite{ahn2010link} to cluster citation links in the bipartite network of papers and their cited sources. Ghosh \textit{et al.} (2011) \cite{Ghosh2011Identifying} have generalised HLC to tripartite graphs.

We did not consider the  line-graph approach because it is not local (due to its use of modularity). 

\subsection{Fuzzification of Hard Clusters} 

\begin{figure}[!b] 
\begin{center} 
\includegraphics[width=3in]{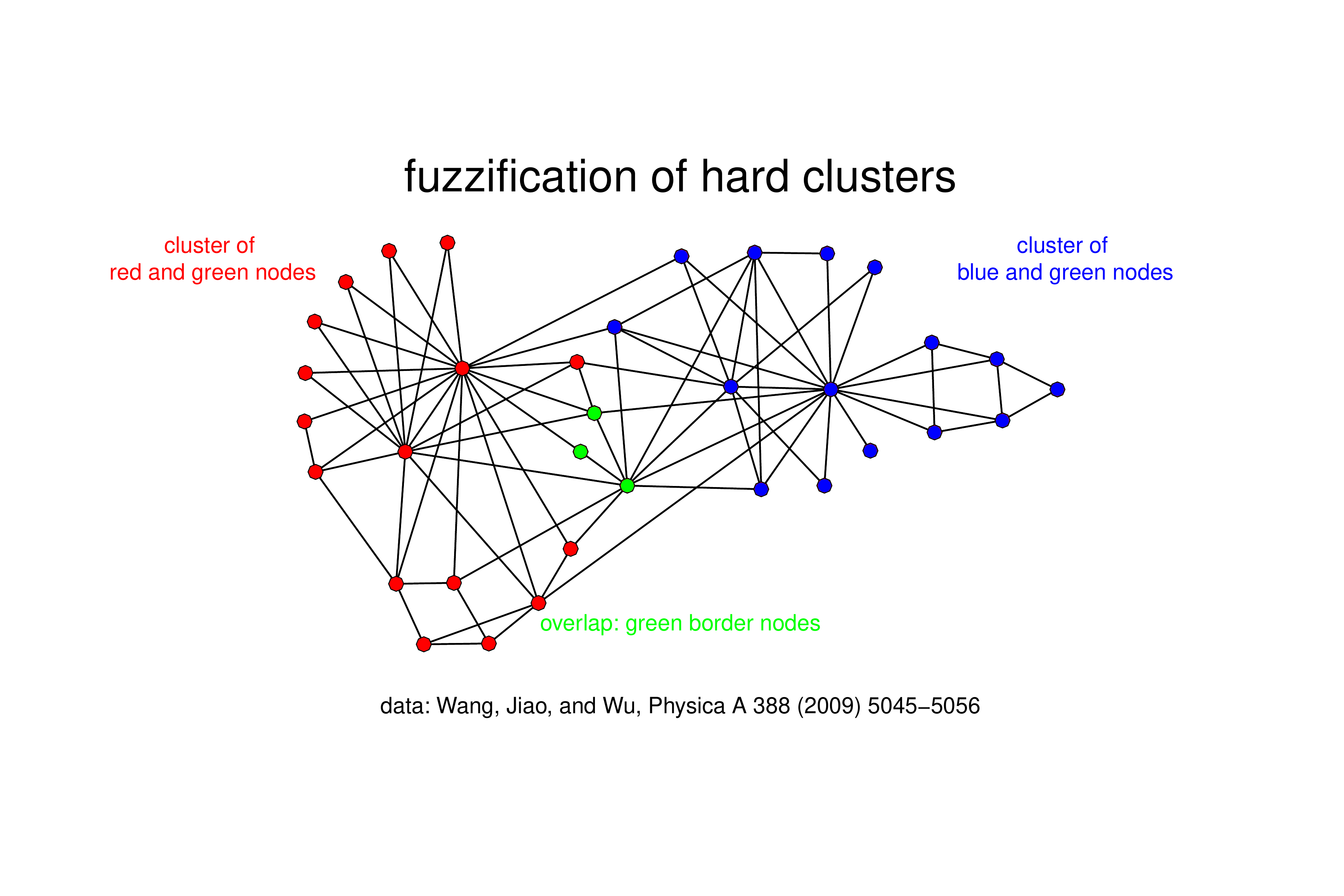} 
\end{center} 
\caption{ {\bf Fuzzification of hard clusters.} Karate club graph with overlapping communities from two hard clusters. } 
\label{Fig-karate-fuzzy} 
\end{figure}

The approach assumes that hard-cluster algorithms validly identify disjoint community cores which just need to be `softened' at the borders. If this is the case, modifying a hard cluster by evaluating the inclusion of its nodes and neighbouring nodes with regard to some metric or fitness is a plausible method for constructing overlapping communities. The fitness balance of a node with respect to a cluster can then be used to decide about its membership and to calculate its membership grade. Thus, we construct fuzzy overlapping communities.

Figure 3 shows the karate-club result Wang \textit{et al.} (2009) \cite{wang2009adjusting} obtained with their implementation of the fuzzification approach, which they applied to two hard clusters.

However, the simplest approach to making hard clusters overlapping and fuzzy is to redefine the membership grades of all nodes with links crossing borders. If some of a node's links end within cluster $C$ and some outside $C$ then its membership in $C$ can be defined as the ratio $\kappa_{in}(C, V_i)/\kappa(V_i)$. We use this definition to calculate fractional grades after we constructed overlapping communities by fitness improvement. Such an approach can be criticised as being inconsistent because fractional memberships are calculated using non-fractional (zero or full) memberships of nodes as obtained by fitness evaluation.

\section{Data}
\subsection{The Paper Network}

We apply the three algorithms to a network of papers in the 2008 volume of six information-science journals with a high proportion of bibliometrics papers (for details of data see reference \cite{havemann2011identification}, papers downloaded from Web of Science).

We start from the affiliation matrix $M$ of the bipartite network of papers and their cited sources. Here we neglect that a few cited sources are also citing papers in the 2008 volume but this minor simplification can be avoided in future analyses. Link clustering is done with $M$ itself, the other two algorithms analyse a bibliographic-coupling network constructed from $M$ as follows. In the network of papers, two nodes (papers) are linked (bibliographically coupled) if they both have at least one cited source in common. To account for different lengths of reference lists we normalise the paper vectors of $M$ to an Euclidean length of one. With this normalisation, the element $a_{ij}$ of matrix $A = M M^T$ equals Salton’s cosine index of bibliographic coupling between paper $i$ and $j$.

The symmetric adjacency matrix $A$ describes a weighted undirected network of bibliographically coupled papers. The main component of the network of 533 information-science papers 2008 (528 articles and five letters) contains 492 papers. Three small components  and 34 isolated papers are of no interest for our cluster experiments.

\begin{figure}[!t] 
\begin{center} 
\includegraphics[width=3in]{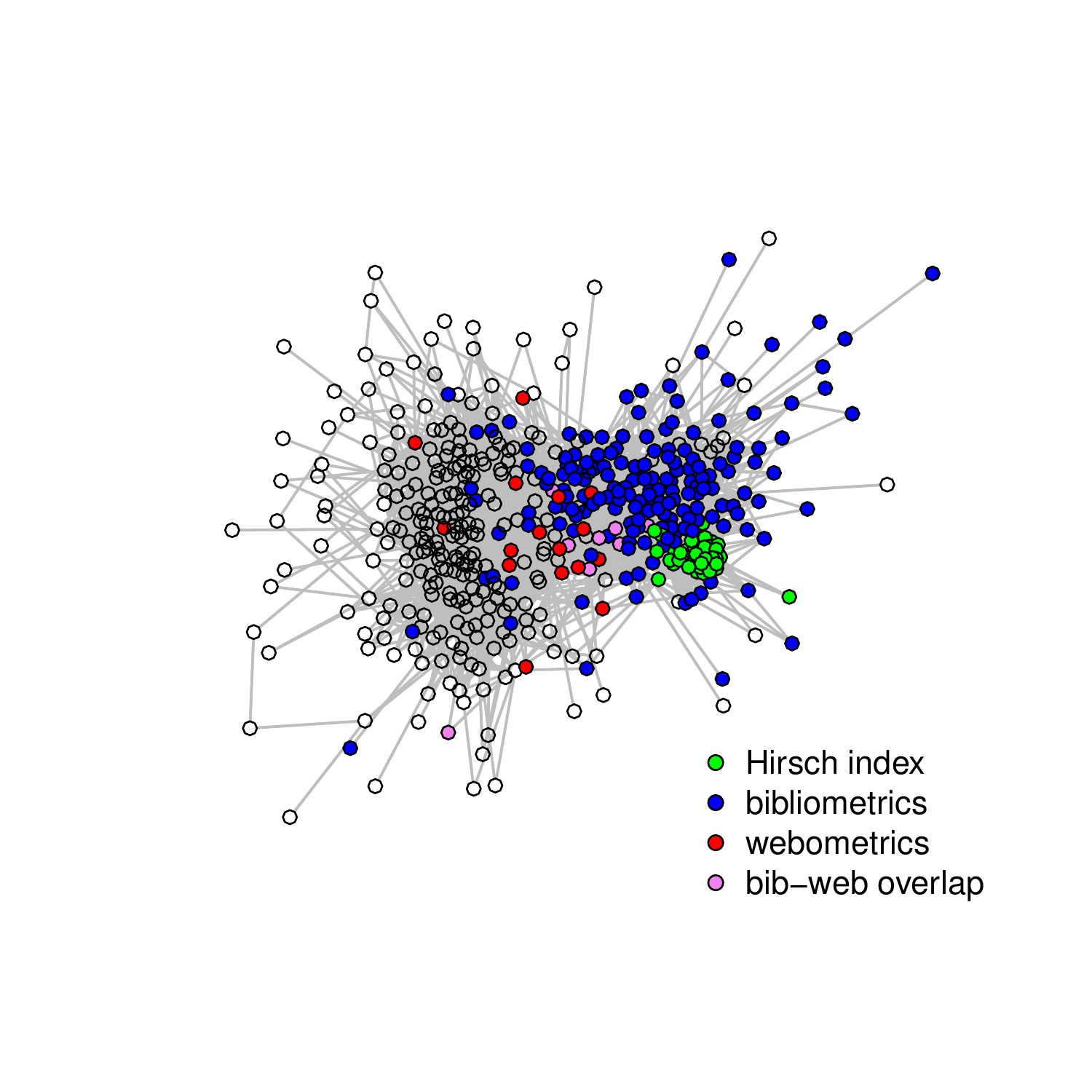} 
\end{center} 
\caption{ {\bf Information science 2008.} Three topics and their overlaps in a network of 492 bibliographically coupled papers. Topics assigned manually to papers by inspection of their keywords, titles and abstracts. The nodes' colours correspond to four sets: (i) green to $h$-index, (ii) blue to bibliometrics without $h$-index and without webometrics, (iii) red to webometrics without bibliometrics, and (iv) violet to the overlap of webometrics and bibliometrics. Transparent nodes are papers dealing with other information-science topics, mainly with information retrieval and information behaviour. } 
\label{Fig-3-topics} 
\end{figure}

\subsection{Three Topics}

For the evaluation of the three algorithms, we used keywords, titles, and abstracts of papers to identify those that belong to three topics, namely $h$-index, bibliometrics, and webometrics. The $h$-index is an indicator for the evaluation of a researcher's performance, which has been proposed by the physicist J.\,E. Hirsch in 2005. Since then, the use of the $h$-index for evaluating individual researchers, proposals for $h$-index derivatives and for $h$-indices of journals or other aggregates of papers have been discussed in the literature. 46 of the 492 papers cite the 2005 paper by Hirsch, which is the most cited source in our sample. The $h$-index is clearly an invention in the field of bibliometrics. About 200 other papers are also addressing bibliometric themes. For the purposes of this evaluation, we excluded analyses of patents from bibliometrics. In a smaller webometrics set, internet activities of (mainly academic) institutions and individuals are analysed.

We first assigned papers to the three topics on the basis of their keywords and subsequently checked the classification by inspecting titles and abstracts. This led to 42 papers assigned to the $h$-index and its derivatives, further 182 bibliometric papers not mentioning the $h$-index in title or abstract, 24 webometric papers, and eight papers in the overlap between webometrics and bibliometrics. In figure 4 we display the graph of the sample of 492 bibliographically coupled papers using the force-directed placement algorithm by Fruchterman and Reingold (as implemented in the \textbf{R}-package \emph{igraph}).\footnote{cf. \url{http://www.r-project.org}}

\section{Fuzzy Natural Communities}
\subsection{MONC Algorithm}

MONC \cite{havemann2011identification} uses ideas from Lancichinetti \textit{et al.} (2009) \cite{lancichinetti2009detecting} but replaces their numerical approach by a faster and more precise parameter-free analytical solution. 

Lancichinetti \textit{et al.} proposed an algorithm which rests on a greedy expansion of natural communities of nodes by local fitness maximisation (LFM algorithm). Communities of different nodes can overlap each other. The size of a natural community of a node depends on resolution $\alpha$. LFM has to be repeated for each relevant resolution level to reveal the hierarchical structure of the network. 
Our parameter-free MONC algorithm exactly calculates resolution levels at which communities change by including a node that improves their fitness. To save further computing time, MONC \textbf{m}erges \textbf{o}verlapping \textbf{n}atural \textbf{c}ommunities when they become identical during the iteration process \cite{havemann2011identification}.

Similar to Lee \textit{et al.} (2010) \cite{lee2010detecting}---who tested a variant of LFM---we found that using cliques as seeds gives better results than starting from single nodes. While Lee \textit{et al.} use maximal cliques (i.e. cliques which are not sub-graphs of other cliques), we optimise clique size by excluding nodes that are only weakly integrated \cite[p. 6]{havemann2011identification}.

From MONC results, we construct \emph{fuzzy natural communities} of nodes i.e. fuzzy sets in which each graph node has a definite membership grade. Each fuzzy natural community represents its seed node's perspective on the whole network. Since the emphasis on local perspectives lets MONC construct many natural communities that are very similar, the fuzzy natural communities are hard-clustered hierarchically using the fuzzy-set Jaccard index as a similarity measure for, e.g., single-linkage clustering. Branches in dendrograms derived from MONC results do not represent disjoint sets of nodes but overlapping fuzzy communities.

\subsection{MONC Post-Processing}

Greedy algorithms which locally maximise a resolution depending fitness (or density) function can reveal hierarchies of overlapping modules. Lancichinetti \textit{et al.} (2009) \cite[pp. 7--9]{lancichinetti2009detecting} have successfully tested their LFM algorithm on a simple benchmark graph with two hierarchical levels.

MONC (like LFM) needs some postprocessing to reveal a graph's hierarchy. We successfully tested the following procedure for detecting a graph's hierarchy from MONC results. We define the grade a node is a member in a community from the resolution level at which it becomes a member (cf. next subsection). With this definition communities become fuzzy sets over the universe of all nodes called \emph{fuzzy natural communities}. Two communities are similar if their fuzzy intersection is large. As a relative measure, we use the fuzzy Jaccard index to define the similarity of two natural communities. Then communities can be clustered by any hard-cluster algorithm to reveal the graph's hierarchy. 

Here we should add a comment. We construct a node's perspective on the whole graph i.e. its natural community as a fuzzy set over the universe of all nodes. We hierarchically cluster the fuzzy sets which is equivalent to node clustering based on a variant of the concept of structural equivalence \cite[p. 86]{fortunato2010community}. Nodes are structurally equivalent if their neighbourhoods are equal, they are structurally similar if their neighbourhoods are similar in some sense. We operationalise structural similarity of two nodes as the fuzzy Jaccard index of their fuzzy natural communities representing their perspectives on the whole graph. Despite the equivalence of our method to the concept of structural similarity of nodes we insist on the definitions given above: we do not cluster nodes but their fuzzy natural communities.

We have also to comment on our earlier claim MONC would reveal the graph's hierarchy automatically, i.e. without postprocessing \cite{havemann2011identification}. MONC's merging of overlapping natural communities can be visualised in a dendrogram. Two communities merge at a resolution level where they become identical sets of nodes. In the case of Zachary's karate club we discussed graph covers on different resolution levels by going through this dendrogram starting from its root. This inspection of a dendrogram of merging communities seemed to confirm our expectation that MONC's community merging is determined by the graph's actual hierarchy \cite[pp. 11--13]{havemann2011identification}. Tests with the information-science network convinced us that this assumption is not true. If two communities merge at a low level of inverse resolution $\gamma=1/\alpha$, they are very similar and are therefore located close to each other in the graph's hierarchy. But merging at low $\gamma$ is not a necessary condition for communities to be similar. There are many very similar communities which merge at high $\gamma$ because of very small differences in membership up to this level. This phenomenon leads to many near-duplicates when we determine modules at one selected resolution level (as we have already found when testing MONC on non-hierarchical benchmark graphs \cite[p. 16]{havemann2011identification}).

\subsection{Grades of Membership}

MONC's greedy expansion of seeds can be discussed in terms of `hosts inviting guests' to their communities. Each node $i$ of the (connected) graph is `invited' to each community $j$ at some level of inverse resolution $\gamma_{ij}$. To construct fuzzy communities with various grades of node membership we propose to define the membership grade of node $i$ in the community of node $j$ as 
\begin{equation} 
\mu_{ij} = \exp(-\gamma_{ij}^2).
\label{def.mu}
\end{equation} 
Using the decreasing exponential function of squared $\gamma_{ij}$ (as in the density function of the normal distribution) ensures that (1) the host is full member in its own community ($\mu_{jj}=1$), (2) `late guests' get lower grades, and (3) the `first guests' get membership grades near one (the function starts from one, its derivation from zero).

We assume that the dendrogram of fuzzy natural communities reflects the graph's hierarchical structure. For each branch we define a community as the fuzzy union of all fuzzy sets of the branch's nodes. This means that all host nodes of the branch are full members of the branch community. This definition ensures the hierarchical order of branches: if two branches unite then their communities are fuzzy subsets of their fuzzy union. Thus, each branch of the dendrogram of fuzzy natural communities, i.e. each vertical line, represents a fuzzy community.

\begin{figure}[!t] 
\begin{center} 
\includegraphics[width=3in]{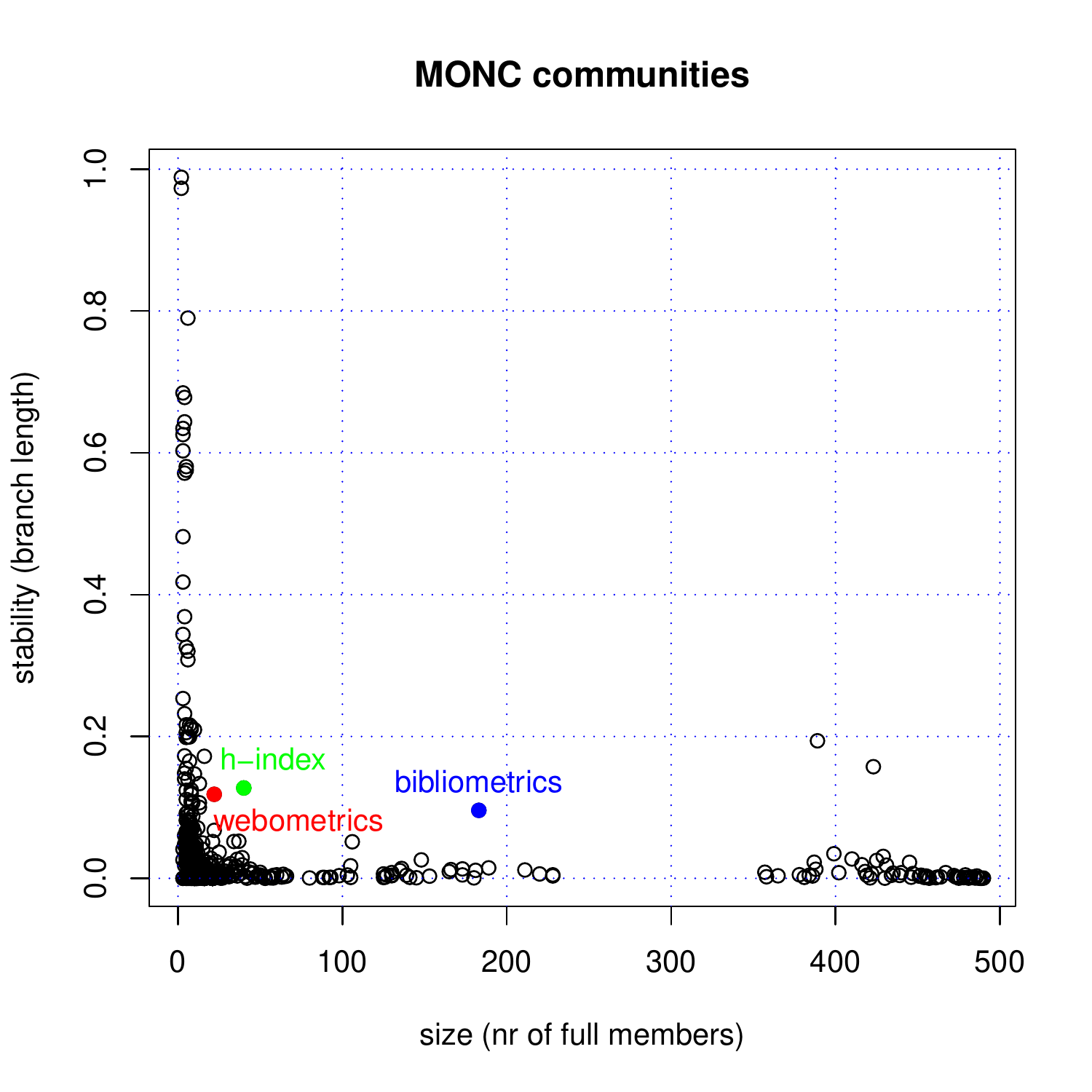} 
\end{center} 
\caption{ {\bf Stability over size of all MONC branch communities.} Stable communities corresponding to our three topics in information-science papers 2008 are marked: bibliometrics (blue), webometrics (red), $h$-index (green). } \label{Fig-MONC-size-stability-plot} 
\end{figure}

\begin{figure}[!p] \begin{center} \includegraphics[width=2.5in]{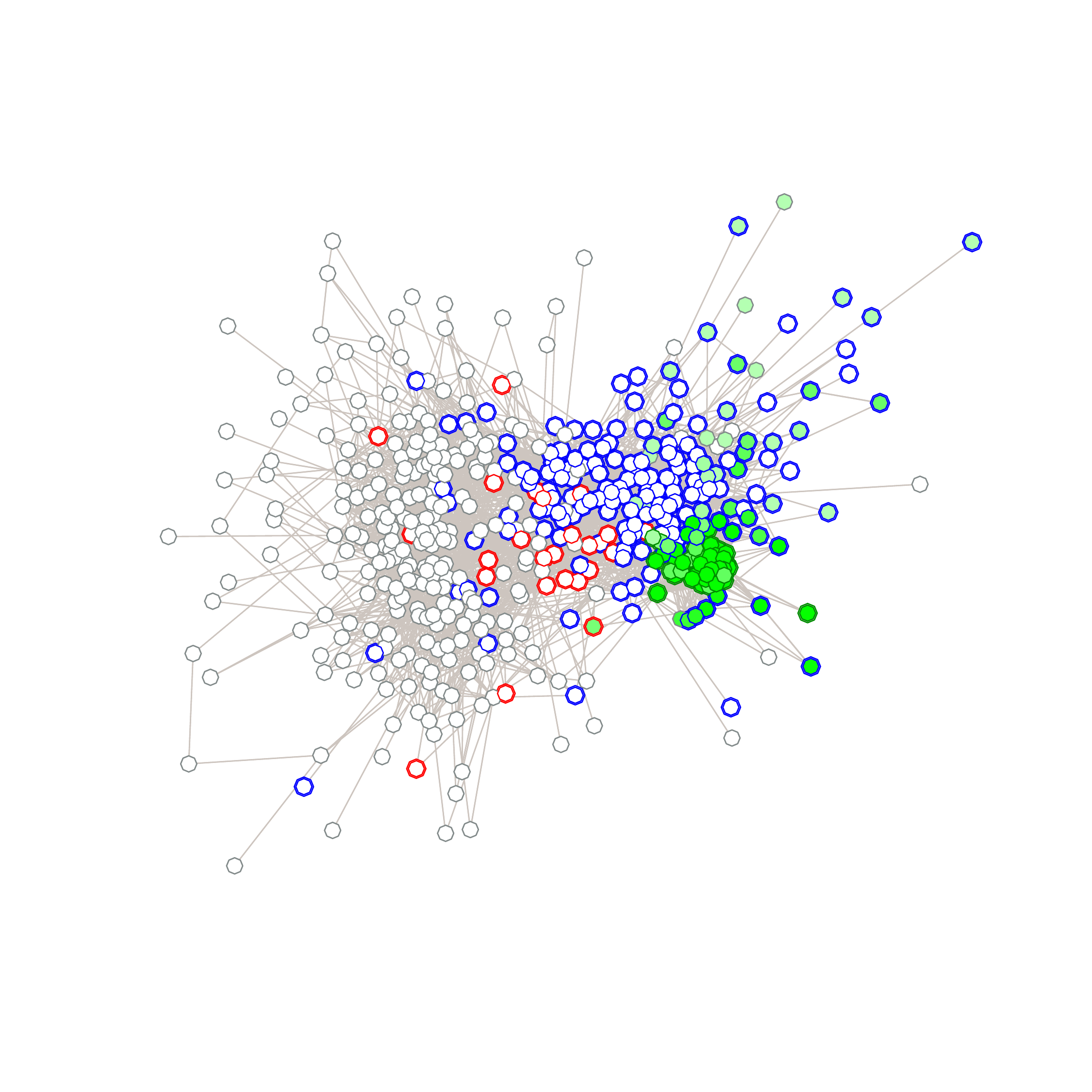} \includegraphics[width=2.5in]{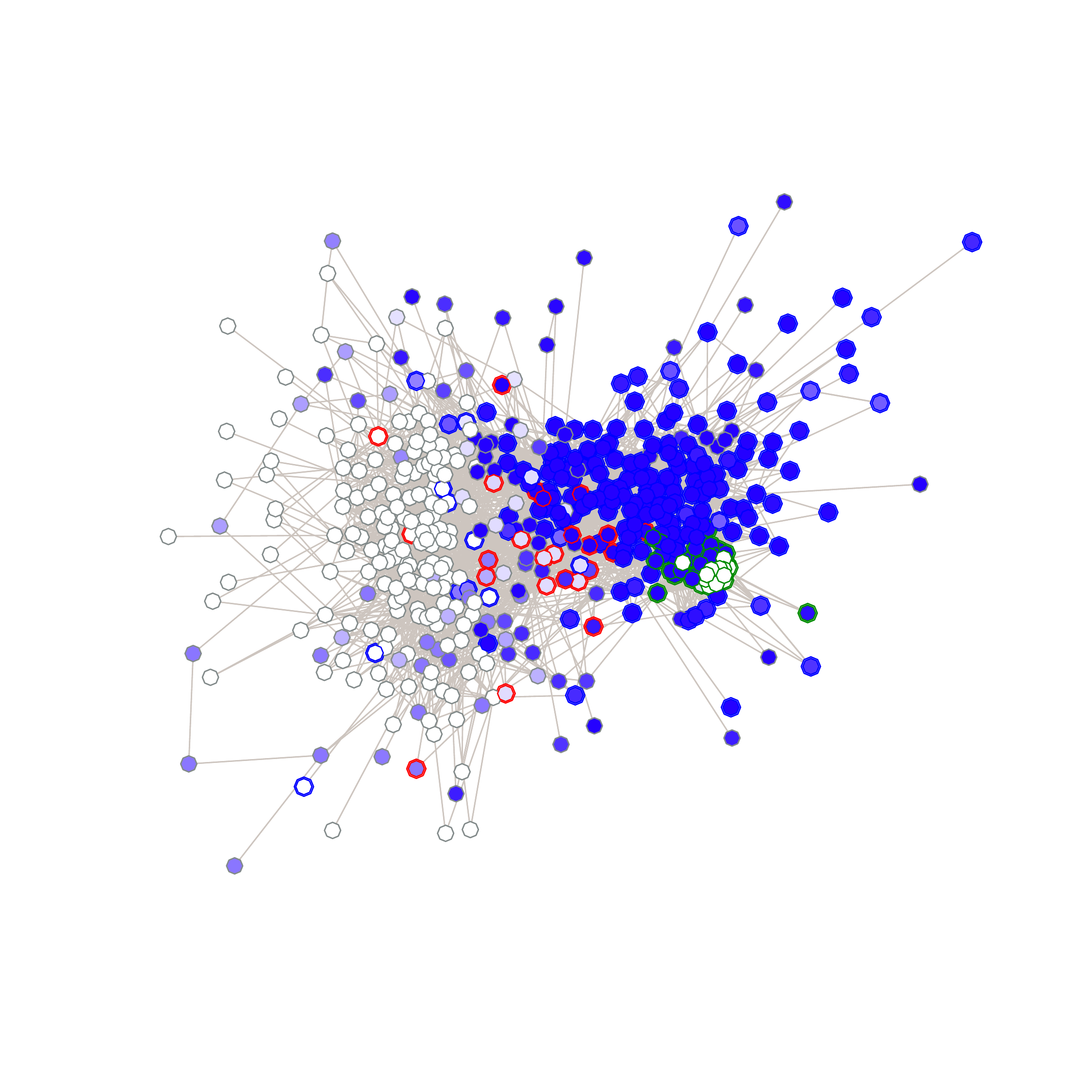} \includegraphics[width=2.5in]{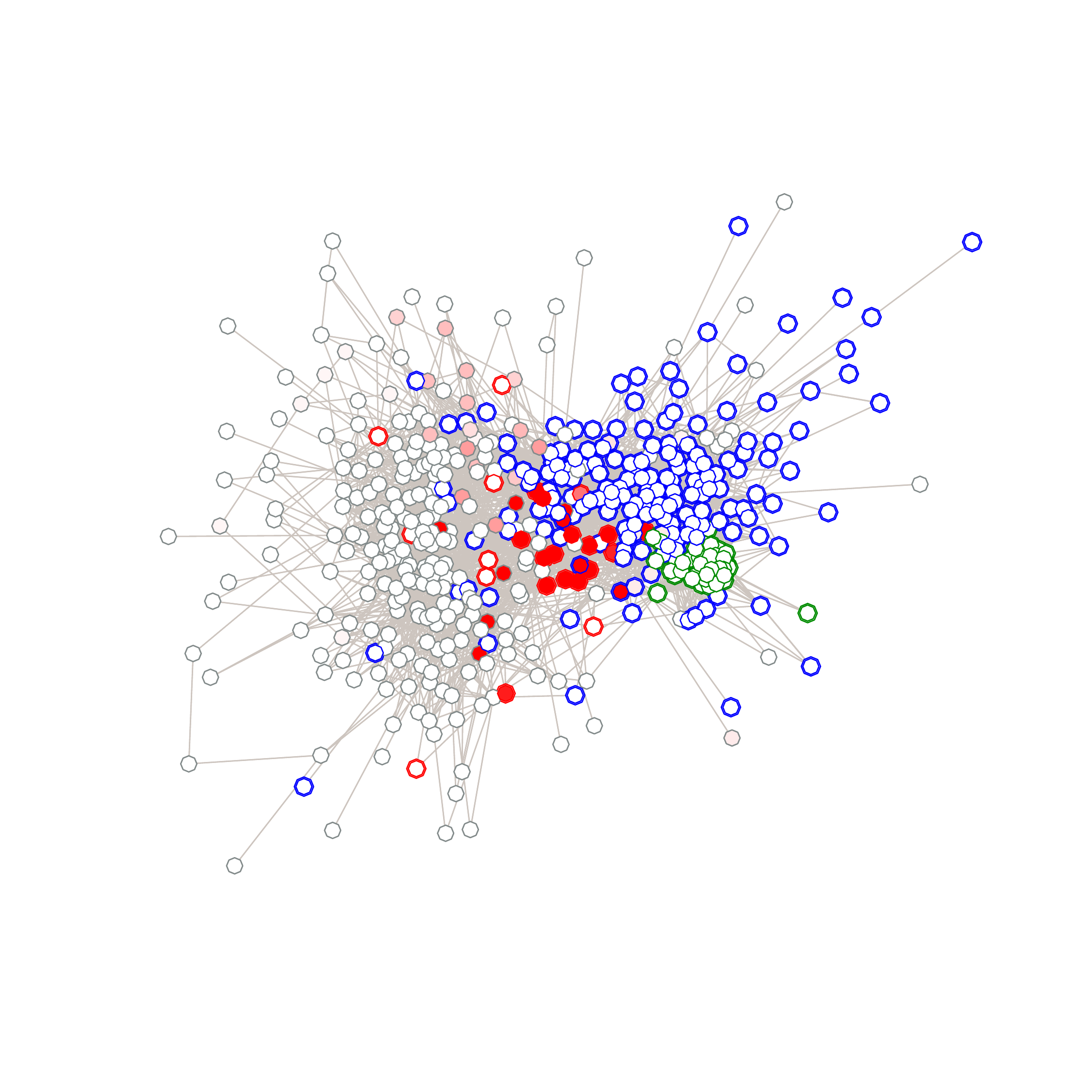}\end{center} 
\caption{ {\bf MONC communities of the three topics} in information-science papers 2008: (a) $h$-index, (b) bibliometrics, (c) webometrics. Saturation of points correlates with membership grade. Colours of circles denote manually determined topics (cf. fig. \ref{Fig-3-topics}).} 
\label{Fig-MONC-topics} 
\end{figure}

\begin{figure}[!p] 
\begin{center} \includegraphics[width=2.5in]{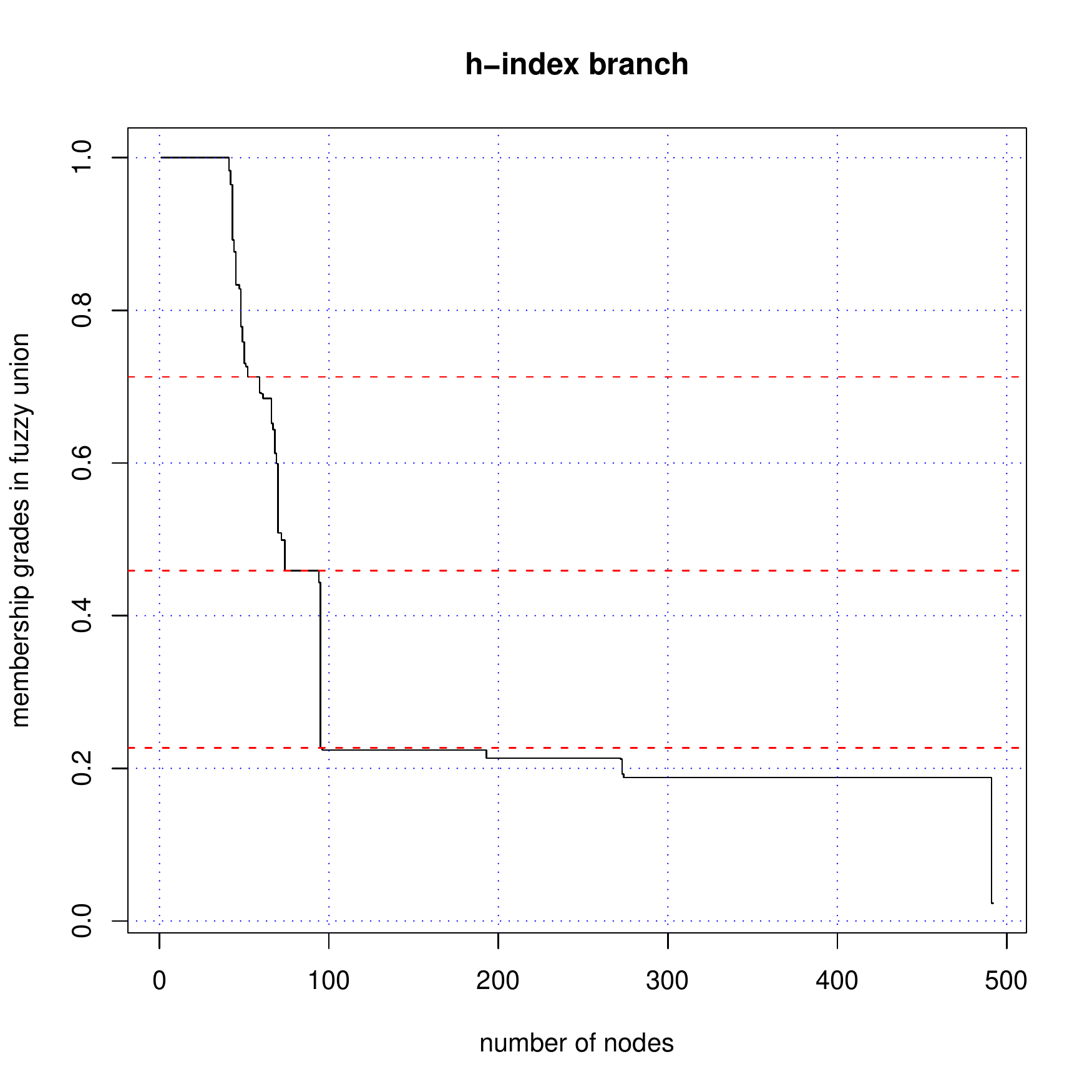} \includegraphics[width=2.5in]{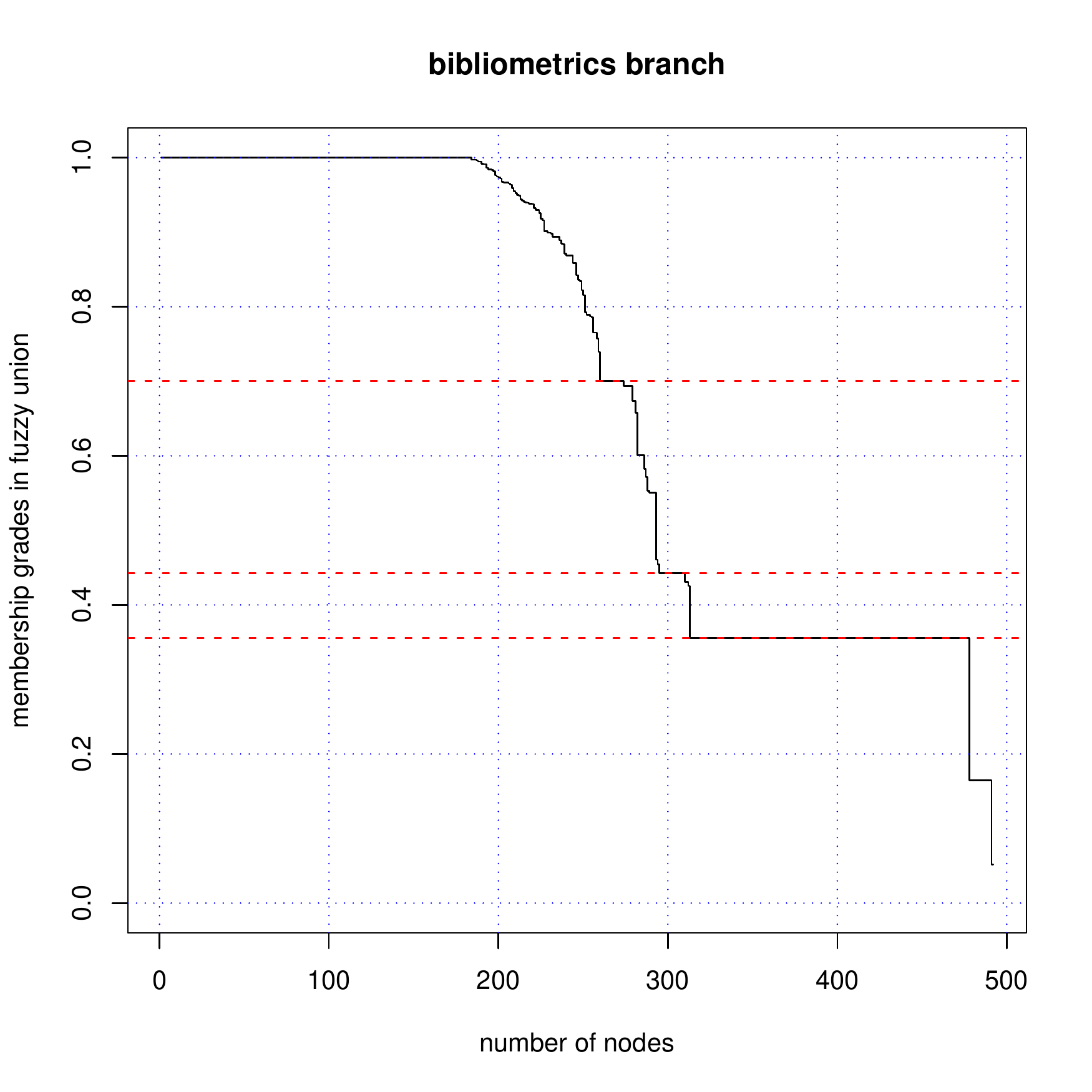} \includegraphics[width=2.5in]{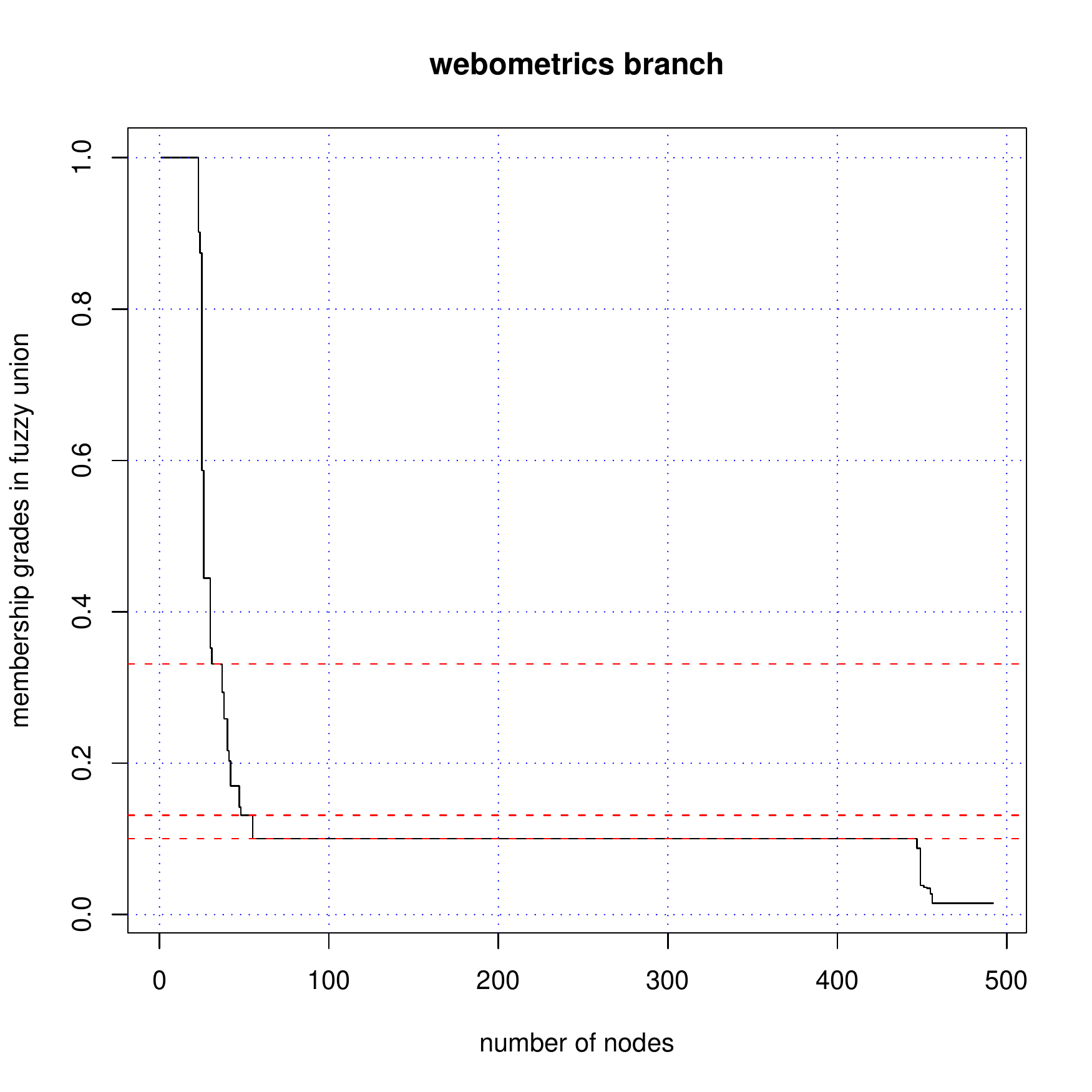} \end{center} \caption{ {\bf Scree plots of MONC communities of the three topics} in information-science papers 2008: (a) $h$-index, (b) bibliometrics, (c) webometrics. Red lines mark possible thresholds.} \label{Fig-MONC-screes} 
\end{figure}

\subsection{MONC Communities of Topics}

For each branch community we plot its stability i.e. its branch's length $s_u - s_d$ over community size, which is estimated by the number of full members (figure \ref{Fig-MONC-size-stability-plot}, cf. also above, p.~\pageref{Communities}). Three of the outliers correspond to our predefined topics, and will be evaluated in comparison with results of the other algorithms (see below, section \textit{Comparison of Algorithms}). The two stable communities with about 400 full members unite bibliometrics, webometrics, information retrieval and some other smaller topics but do not include a set of less central graph nodes.

Figure \ref{Fig-MONC-topics} shows the relationship between MONC communities corresponding to the three topics and the manually determined paper sets of topics. All grades below a threshold are set to zero. We derive the thresholds from scree plots of membership grades (figure \ref{Fig-MONC-screes}). These plots show that at some critical membership grade the node sets of each branch inflate to nearly the whole graph. We argue that this inflation marks the border of a community of a branch. We set the grade's threshold on a value that cuts the scree at the last steepest gradient before inflation ($\mu_{\mathrm{thr}} = .229,\, .355,\, .1$ for $h$-index, bibliometrics, and webometrics, respectively).

\section{Hierarchical Clustering  \\of Citation Links}
\subsection{HLC Algorithm \\on Bipartite Graphs}

We consider the bipartite network of papers and cited sources. Citation links between the two types of nodes can be hard-clustered, which leads to induced overlapping communities of papers (and also to communities of sources which, however, are not analysed here). The membership grade of a pa­per to a thematic community equals the fraction of its citation links belonging to the corresponding link cluster.

Links can be seen as similar if the neighbourhoods of their nodes overlap to a high degree. Thus, the Jaccard index of these neighbourhoods can be used as a similarity measure (cf. Ahn \textit{et al.}, 2010 \cite[eq.~2, p.~5]{ahn2010link}). We discuss the definition of similarity between links in a bipartite graph in terms of papers and cited sources. The neighbourhood of a paper $p_i$ is the set of its references $R_i$, the neighbourhood of a cited source $s_i$ is the set of papers $C_i$ citing it. The neighbourhood $N_i$ of citation link $i$ is then the union of these disjoint\footnote{We neglect that a few cited sources are also citing papers, cf. above.} sets: $N_i = R_i \cup C_i$. The size of the intersect of two link neighbourhoods is given by 
\begin{equation}
|N_i  \cap N_j |= |C_i \cap C_j| + |R_i \cap R_j| 
\end{equation} 
and the size of their union by 
\begin{equation} 
|N_i \cup N_j |= |C_i \cup C_j| + |R_i \cup R_j|. 
\end{equation} 
The distance metrics used for clustering is then 
\begin{equation} 
d_{ij} = 1 - \frac{|C_i \cap C_j| + |R_i \cap R_j|}{|C_i \cup C_j| + |R_i \cup R_j|}. \label{eq-HLC-dist} 
\end{equation}

Ahn \textit{et al.} calculate similarities only for link pairs which have a node in common because they ``expect'' disconnected link pairs to be less similar then pairs connected over a node \cite[p. 5]{ahn2010link}. Since counterexamples disproving this assumption can be constructed, we decided to calculate similarities for all pairs of nodes. Such a procedure uses more information but  is also more time-consuming.

\begin{figure}[!t] 
\begin{center} 
\includegraphics[width=3in]{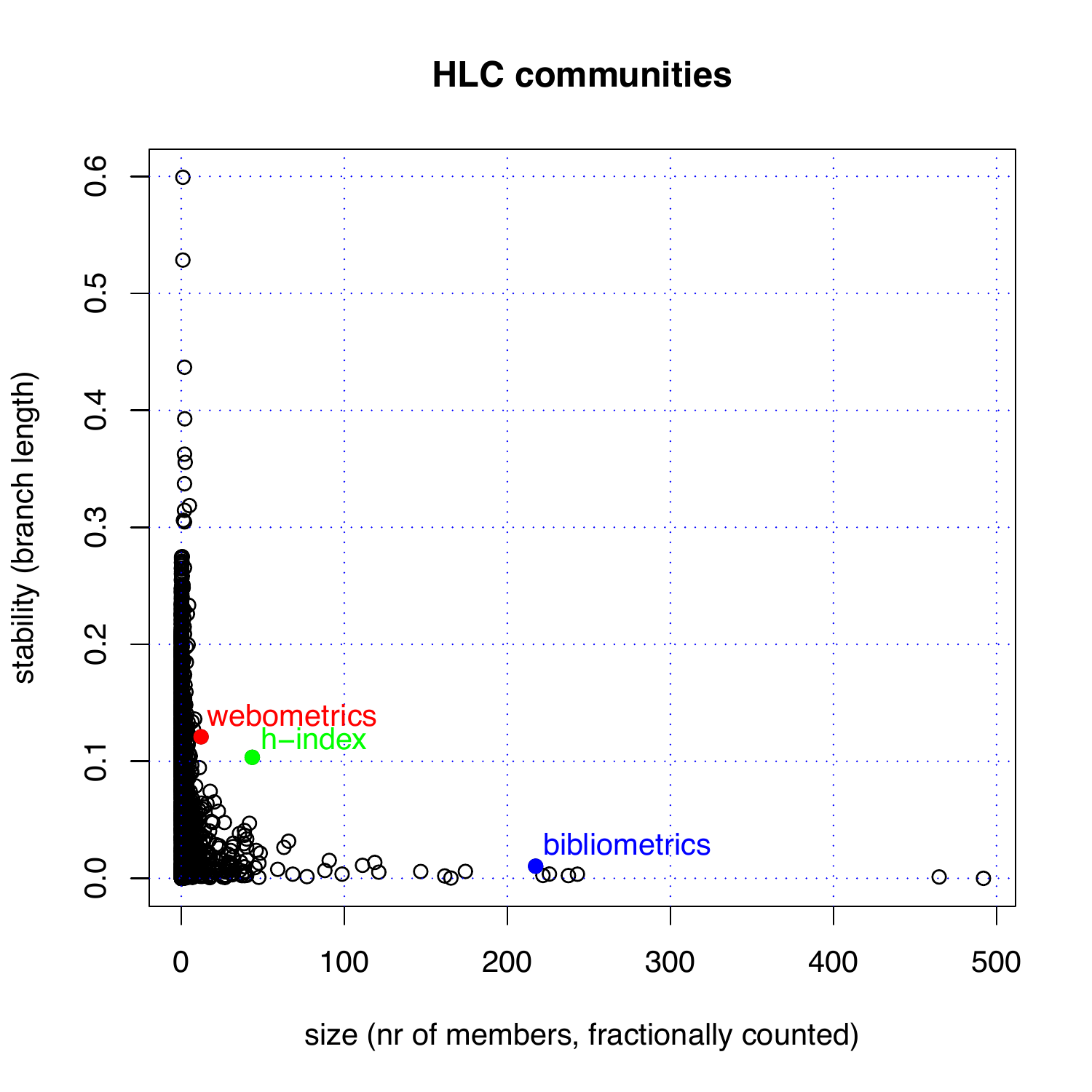} 
\end{center} 
\caption{ {\bf Stability over size of all 5004 HLC branch communities.} Stable communities corresponding to our three topics in information-science papers 2008 are marked: bibliometrics as blue, webometrics as red, and $h$-index as green point, respectively.} 
\label{Fig-HLC-size-stability-plot} 
\end{figure}

\begin{figure}[!p] 
\begin{center} 
\includegraphics[width=2.5in]{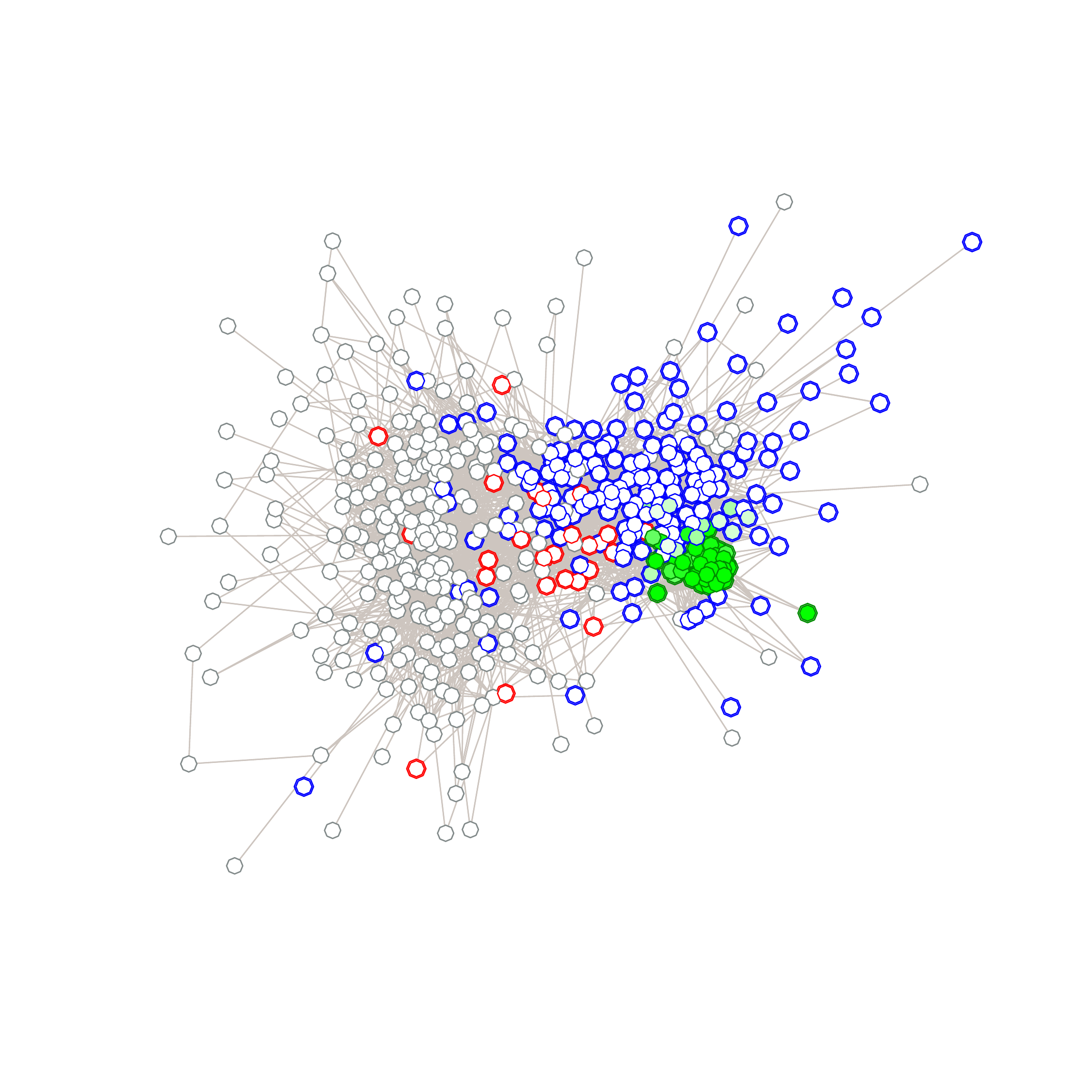} \includegraphics[width=2.5in]{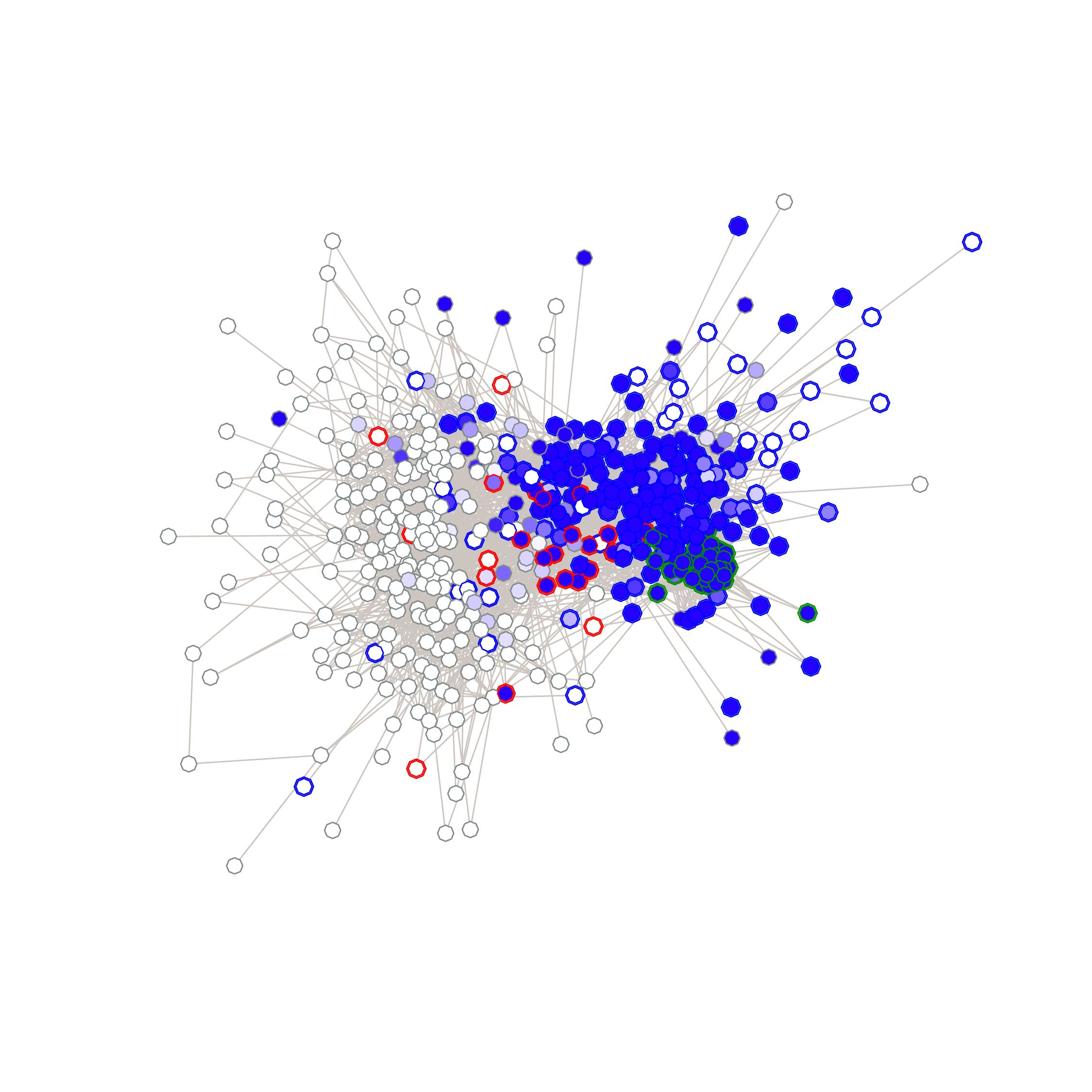} \includegraphics[width=2.5in]{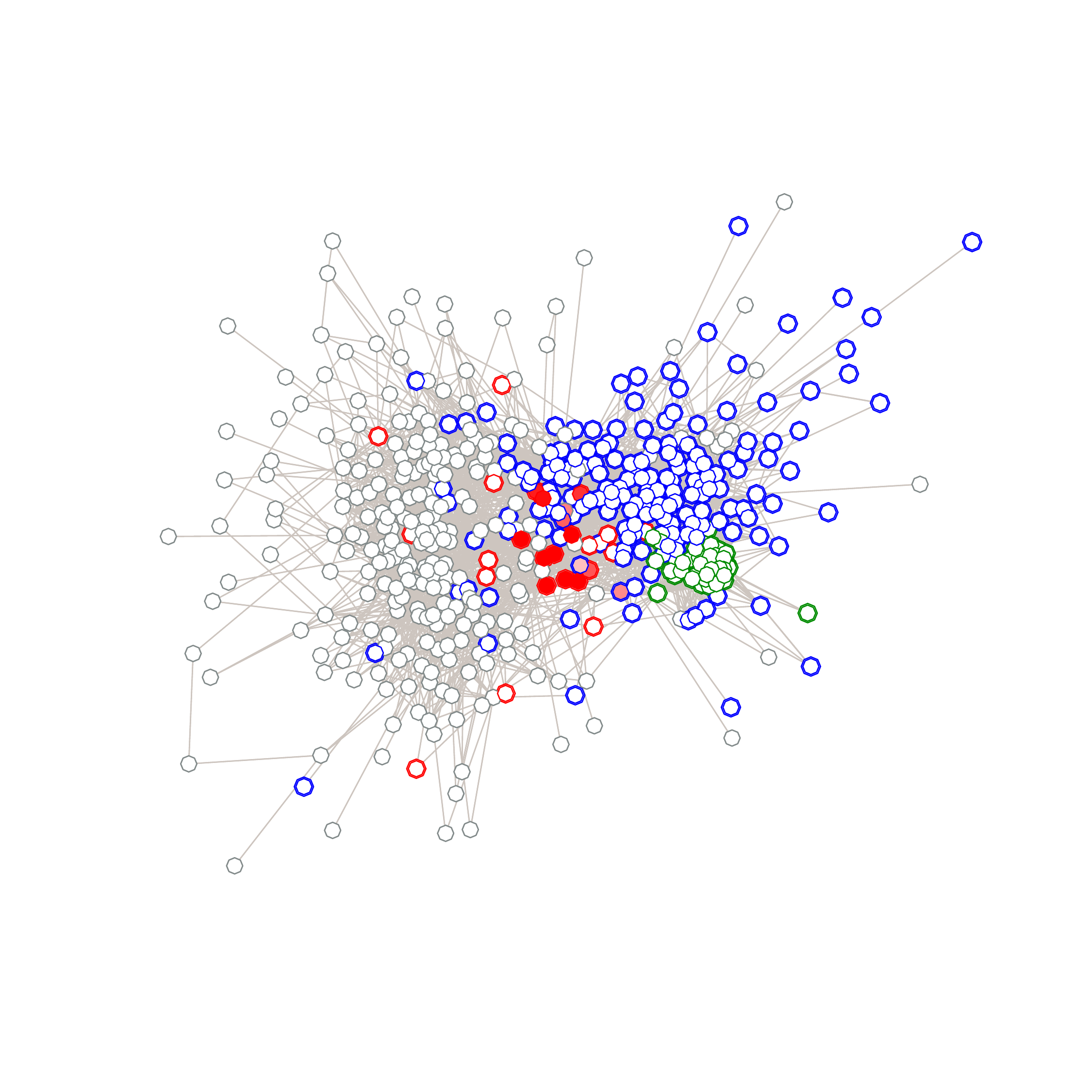} 
\end{center} 
\caption{ {\bf HLC communities of the three topics} in information-science papers 2008: (a) $h$-index, (b) bibliometrics, (c) webometrics. Saturation of points correlates with membership grade. Colours of circles denote manually determined topics (cf. fig. \ref{Fig-3-topics}).} \label{Fig-HLC-topics} 
\end{figure}

Hard clustering of links can be done with any hierarchical clustering method. We tested four standard methods. The dendrograms of Ward and average clustering of citation links seem to reflect the graph's hierarchy more adequately than those of single-linkage and complete-linkage clustering. The latter two methods impose too low or too high restrictions, respectively, on finding clusters.

\subsection{HLC Communities of Topics}

For all pairs of citation links from the 492 citing papers to all sources we determine link similarities. Pairs of citation links to sources cited only once have zero distance within their reference list (cf. equation \ref{eq-HLC-dist}). They are clustered at zero-distance level with one another. Next, these zero-distance clusters are joined with links to the least cited source in their reference list. This allows us to restrict clustering to all $m = 5005$ citation links to sources which are cited more then once.

We applied the average-clustering method to this set. The corresponding dendrogram does not give a clear picture of the graph's hierarchy unless we re-parametrise the distance axis. We choose $d \to d^{\log_2 m}$ to de-skew distances $d$. Using these rescaled data, we plot branch length over community size to find relatively stable and large communities (figure~\ref{Fig-HLC-size-stability-plot}). We measure community size by the sum of fractional membership grades of papers attached to the clustered citation links. Like in the MONC case, we find our three topics as exceptional points in the plot.

Figure \ref{Fig-HLC-topics} shows the graph of 492 papers coloured proportional to their membership grade in the three topics, respectively. 

\section{Fuzzification of \\Hard Clusters}
\subsection{Fuzzification Algorithm}
\begin{figure}[!t] 
\begin{center} 
\includegraphics[width=3in]{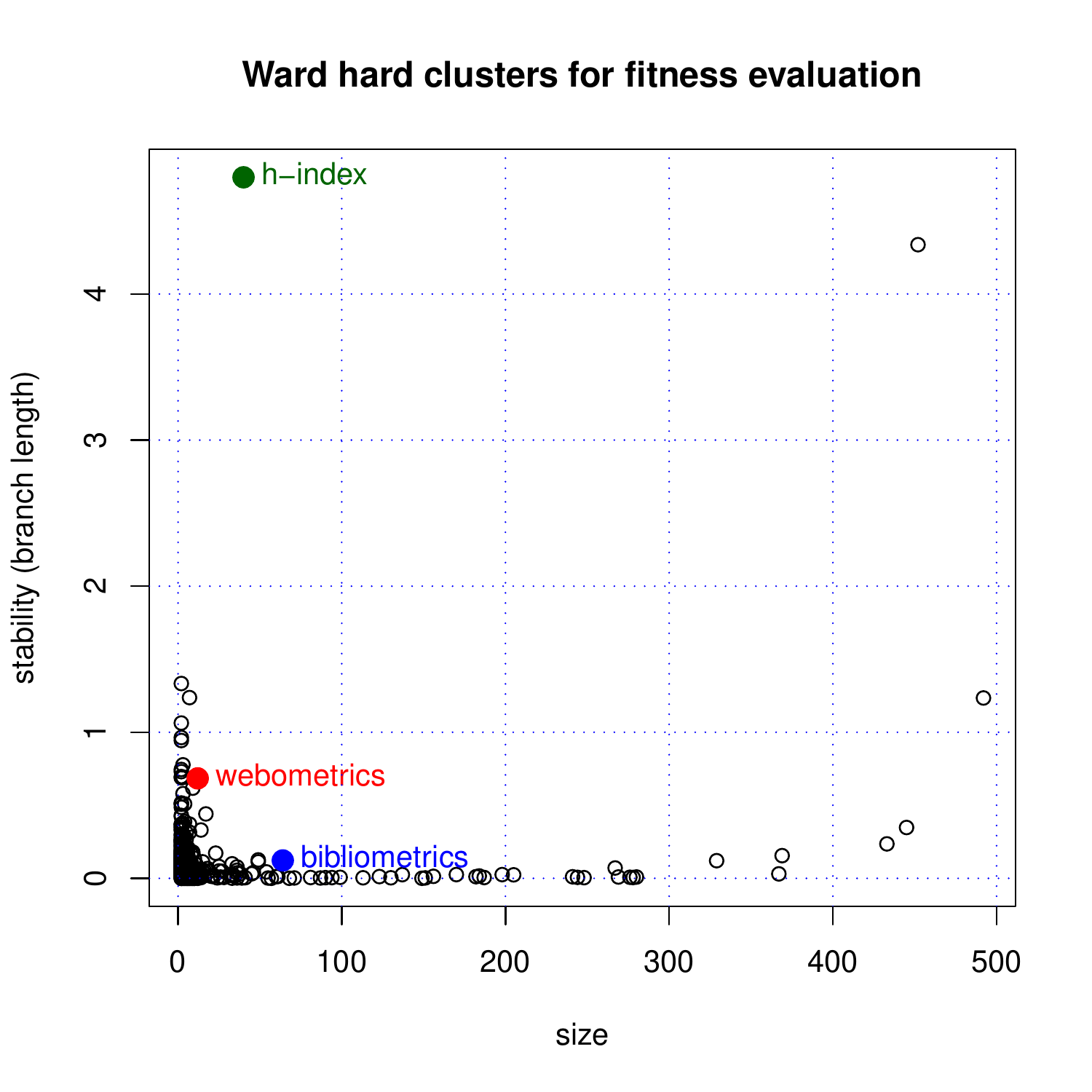} 
\end{center} 
\caption{ {\bf Stability over size of  branch communities.} Stable Ward clusters corresponding to our three topics in information-science papers 2008 are marked: bibliometrics as blue, webometrics as red, and $h$-index as green point, respectively.} 
\label{Fig-fuzzy-size-stability-plot} 
\end{figure}

\begin{figure}[!p] \begin{center} 
\includegraphics[width=2.5in]{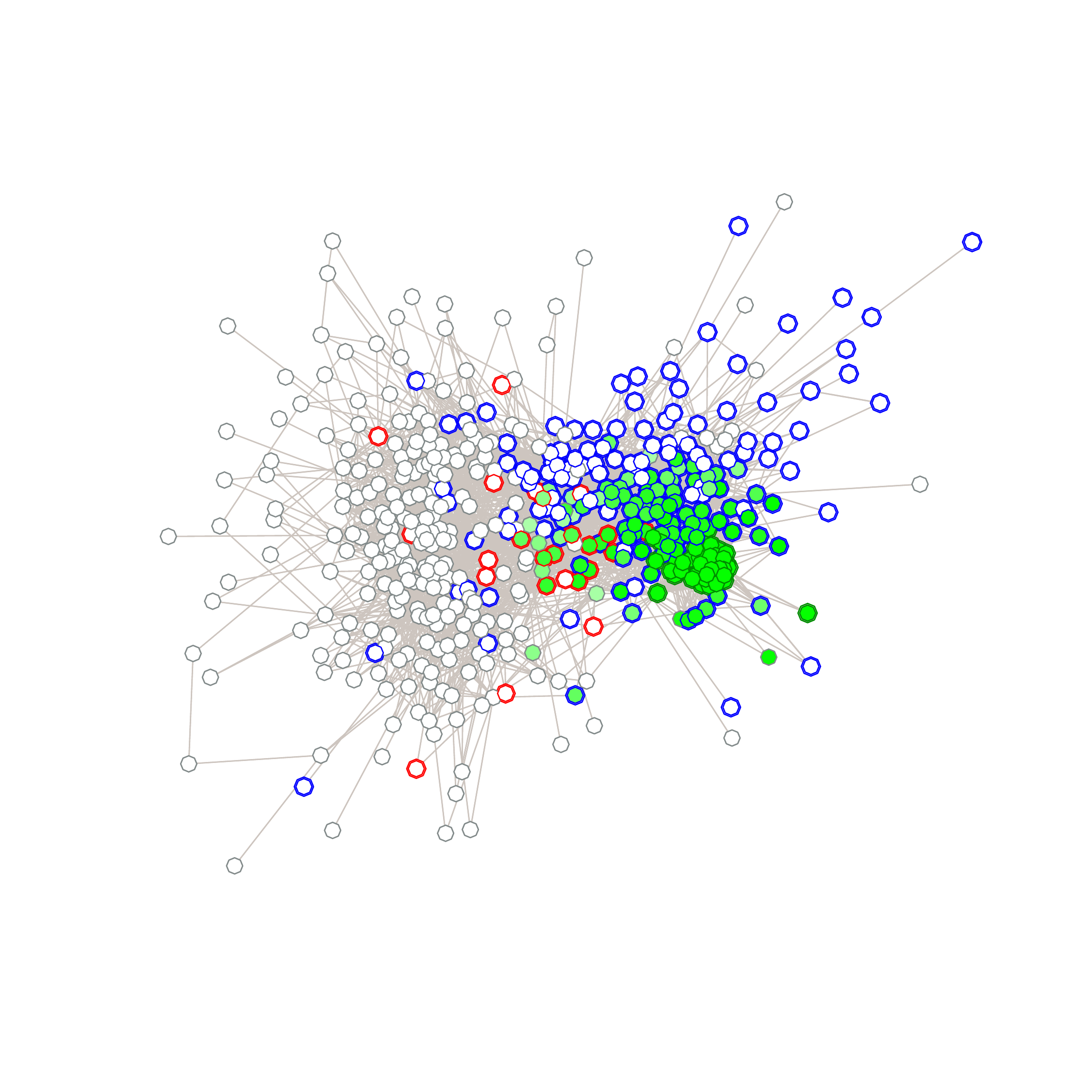} \includegraphics[width=2.5in]{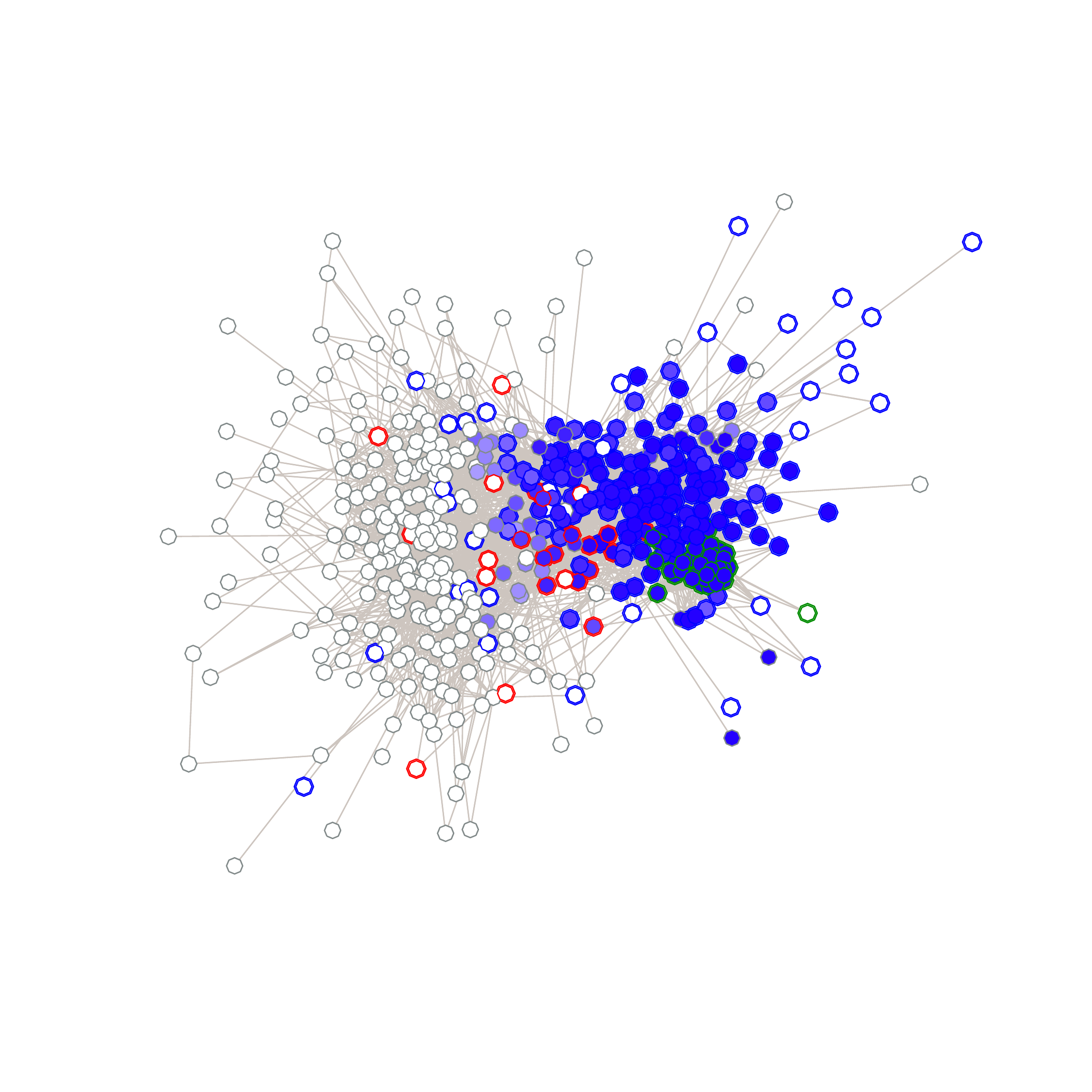} \includegraphics[width=2.5in]{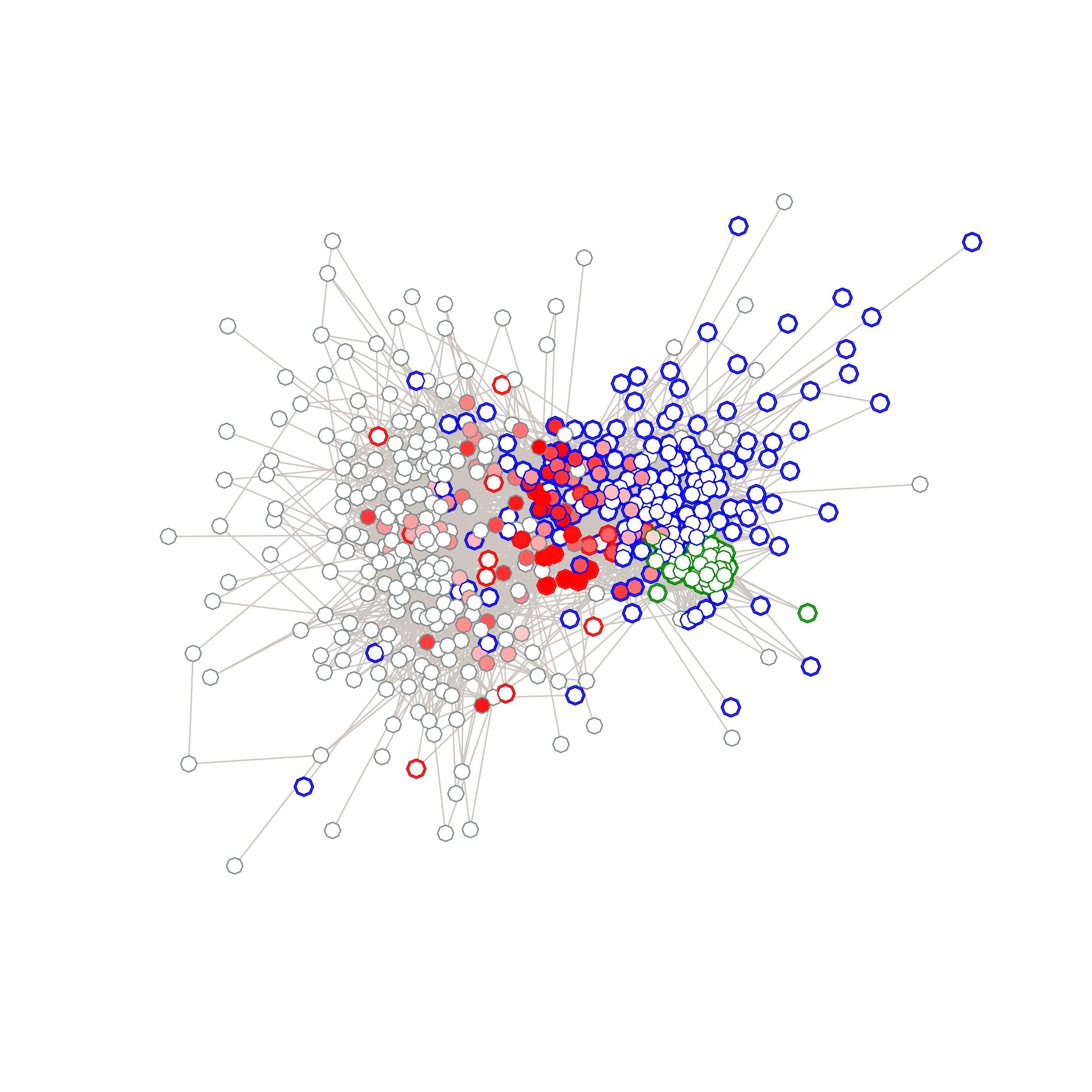} \end{center} 
\caption{ {\bf Fuzzy communities of the three topics} in information-science papers 2008: (a) $h$-index, (b) bibliometrics, (c) webometrics. Saturation of points correlates with membership grade. Colours of circles denote manually determined topics (cf. fig. \ref{Fig-3-topics}).} \label{Fig-fuzzy-topics} 
\end{figure}
The fuzzification approach assumes that hard cluster algorithms validly identify disjoint community cores which just need to be `softened' at the borders. We implemented an algorithm that evaluates border nodes of each hard cluster with regard to their connectiveness with it. Border nodes have edges crossing the cluster's border and can be located inside or outside the cluster.

The algorithm uses an evaluation metric that is based on the fitness function defined by Lancichinetti \textit{et al.} (2009) \cite{lancichinetti2009detecting,wang2009adjusting} (see above, equation \ref{def.fitness}, page \pageref{def.fitness}). For each border node of a cluster we calculate the clusters's fitness with and without this node. The fitness balance of a node with respect to a cluster determines its membership. Negative balance means exclusion from the cluster. We evaluate all border nodes of a cluster without changing it during the evaluation (in contrast to the greedy LFM algorithm which updates the community after deciding about a node's membership). The fitness-inherent resolution para­meter controls the extent of the overlap, where lower values cause a wider area to be considered for the inclusion into the former hard cluster. While MONC uses resolution levels to calculate membership grades, the fitness-inherent parameter is arbitrary here. Thus, we didn't apply it in our comparison and set $\alpha=1$. In a second step the crisp overlapping communities are made fuzzy. The fractional membership grade of a node could be defined using its fitness balance as input but this did not lead to fuzzy communities that match the three predefined topics. Hence, we used a definition that ignores the value of (positive) fitness balances: The membership grade $\mu_i(C)$ of vertex $V_i$ in community $C$ is
\begin{equation}
\mu_i(C) =\kappa_{in}(C, V_i)/\kappa(V_i),
\label{def.fuzzy.mu}
\end{equation} 
where $\kappa_{in}(C, V_i)$ is the sum of weights of edges between vertex $V_i$ and vertices in $C$ and $\kappa(V_i)$ the sum of all its edge weights.

\subsection{Fuzzy Communities of Topics}

We applied standard Ward and average clustering on the network of $n=492$ bibliographically coupled information-science papers. Complete and single linkage failed to provide acceptable results as can be already deduced from the dendrograms. Average clustering also results in a dendrogram which is not easy to interpret. The Ward dendrogram shows a very stable and clear $h$-index cluster which is united with the rest of the graph in the last merging step (cf. figure \ref{Fig-fuzzy-size-stability-plot}). Fitness-based optimisation with resolution $\alpha = 1$ enlarges this cluster extremely and lowers precision without gain in recall with respect to the set of manually selected $h$-index papers. Thus, fitness maximisation is not a successful strategy for this topic that has been well matched (by e.g. Ward) clustering already and is highly connected to its network environment.

If we omit fitness maximisation and only calculate fractional membership grades according to equation \ref{def.fuzzy.mu} the result is not better. Many
external border nodes become partial members of the fuzzy $h$-index community.

On the other hand, the hard bibliometrics cluster is much smaller than expected and needs fitness maximisation or at least fractional membership grades to match the topic. Figure \ref{Fig-fuzzy-topics} shows how the fitness-optimised and fuzzyfied Ward clusters of the three topics fit the topics as paper sets. 

\section{Comparison of Algorithms}
\subsection{Comparison of the \\Identified Communities}

To compare the results we calculate fuzzy Salton's cosine of manually defined topics with fuzzy communities identified by the three algorithms considered. In addition, table \ref{tab:label} gives values of fuzzy $k_\mathrm{in}$ and $k_\mathrm{out}$, the geometric mean of which equals the cosine. 
% FH new:
Table \ref{tab:between algorithms} shows how fuzzy communities constructed by the algorithms overlap each other.
In table \ref{tab:k_in.k_out} we list the fuzzy internal and external degrees, $k_\mathrm{in}(C)$ and $k_\mathrm{out}$ (cf. equations \ref{fuzzy-k_in} and \ref{fuzzy-k_out}, p. \pageref{fuzzy-k_in}), together with their ratio $k_\mathrm{in}(C)/k_\mathrm{out}(C)$ for each fuzzy topic community constructed by the three algorithms. All fuzzy communities are communities in the weak sense. The ratio can be interpreted as a measure of `communityness'.
 
The assumption that hard clusters can be improved by fitness-based optimisation
and fuzzification could not be validated with Ward clusters as input. While the optimised and fuzzified bibliometrics cluster gained slightly better similarity results than the other two algorithms, the clearly identified $h$-index hard-cluster did not improve because both optimisation and fuzzification included too many nodes which were related but were not assigned to the topic. The fitness-inherent resolution parameter could improve similarity values but would have to be choosen differently for different clusters---a procedure which cannot be applied when target topics are not known in advance. The fact, that fuzzification results in an $h$-index community with best ratio $k_\mathrm{in}(C)/k_\mathrm{out}(C)$ should be interpreted with care. It only means, that the algorithm finds a big cluster which is relatively separated from the rest of the graph. It (partly) includes many papers which do not refer to the $h$-index.

\begin{table}[!b]
\caption{
 \bf{Topic matches by algorithms}}
\begin{center}
 \begin{tabular}{rrrr}
topic & MONC  &  HLC & fuzzy \\
\hline
\textit{$h$-index} &.71 &.93 & .59\\
precision& .56 & .91 & .35\\
recall & .89 & .95 & 1.00 \\
\hline
\textit{bibliometrics} &.79 & .82 & .83\\
precision&.72 & .83& .87\\
recall& .86& .81& .80\\
\hline
\textit{webometrics} &.58 &.60 & .46\\
precision& .53& .85& .45\\
recall&.64 &.43 & .47\\
\hline
\textit{bib-web overlap} & .46 & .29  &  .30\\
precision & .34 & .24 & .14\\
recall & .64 & .36 & .65\\\hline
 \end{tabular}
\end{center}
\begin{flushleft}Fuzzy cosine indices, precision, and recall of paper sets and fuzzy communities (and of bibliometrics-webometrics overlap) found by the three algorithms 
\end{flushleft}
 \label{tab:label}
  \end{table}

\begin{table}[!b]
\caption{
 \bf{Community matching \textit{between} algorithms}}
\begin{center}
 \begin{tabular}{rrrr}
      & MONC  &  HLC & fuzzy \\
topic & HLC & fuzzy & MONC \\
\hline
\textit{$h$-index}     & .73 & .60 & .62\\
%\hline
\textit{bibliometrics} & .76 &  .84 & .78\\
%\hline
\textit{webometrics}   & .63 & .46 & .55\\
%\hline
\textit{bib-web overlap} & .51 & .41 & .43 \\
\hline
 \end{tabular}
\end{center}
\begin{flushleft}Fuzzy cosine indices of fuzzy communities (and of bibliometrics-webometrics overlap) found by the three algorithms 
\end{flushleft}
 \label{tab:between algorithms}
  \end{table}

\begin{table}[!b]
\caption{
 \bf{Fuzzy $k_\mathrm{in}/k_\mathrm{out}$ of  communities }}
\begin{center}
 \begin{tabular}{rrrrr}
 $C$& variable & MONC  &  HLC & fuzzy \\
\hline
\textit{$h$-index} &$k_\mathrm{in}/k_\mathrm{out}$ & 5.97 & 7.41 & 9.70\\
&$k_\mathrm{in}$& 244.65 & 245.66 & 352.21\\
&$k_\mathrm{out}$ & 40.98 & 33.17 &  36.31\\
\hline
\textit{biblio-}& $k_\mathrm{in}/k_\mathrm{out}$& 3.41 & 19.03 & 15.37\\
\textit{metrics}&$k_\mathrm{in}$& 314.23 & 466.97 & 456.97\\
&$k_\mathrm{out}$& 92.03 & 24.54 & 29.74\\
\hline
\textit{webo-}& $k_\mathrm{in}/k_\mathrm{out}$& 1.43 & 1.21 &  1.32\\
\textit{metrics}&$k_\mathrm{in}$& 21.04  & 10.74 &  45.01\\
&$k_\mathrm{out}$& 14.71 & 8.85 &  34.19\\
\hline

 \end{tabular}
\end{center}
\begin{flushleft}The ratio $k_\mathrm{in}(C)/k_\mathrm{out}(C)$, $k_\mathrm{in}(C)$, and $k_\mathrm{out}(C)$ of fuzzy communities found by the three algorithms 
\end{flushleft}
 \label{tab:k_in.k_out}
  \end{table}

Hierarchical clustering of citation links gave better results than MONC. Link clustering classifies $h$-index as a bibliometric topic whereas MONC only includes some $h$-index papers into bibliometrics. Fuzzy cosines of HLC communities and manually selected topics are always better than the corresponding MONC values (s. table \ref{tab:label}).

\subsection{Assumptions Used}

All three algorithms implemented by us are based on the assumptions that a graph's communities (1) are best determined locally, (2) overlap each other, (3) are best described by fractional membership grades, and (4) form a hierarchy. 
We used these four assumptions as criteria in our selection of approaches to community detection. Nonetheless, with respect to all four criteria there are differences between the selected algorithms. Another criterion was that results should not---at least not strongly---depend on arbitrary parameters. Furthermore, each of the algorithms is based on specific assumptions, which we already mentioned in the respective sections of this paper.

\textbf{The fuzzification procedure} based on hard clusters whose fitness is improved assumes that the hierarchical cluster algorithm delivers essentially valid but improvable hard clusters. Our results do not confirm the improvement using standard fitness measures.
For the membership grades we have not found a local, consistent, and realistic definition. With respect to its input data, this procedure---like MONC but unlike HLC---assumes that a network of scholarly papers weighted with paper similarity (based on references and/or text) can be used to identify hierarchical thematic structures.

\textbf{Hierarchical link clustering:} Paper networks are projections of bipartite graphs and thus do not use the full information content of the raw data. Hierarchical link clustering (HLC) rests on a broader information basis when applied to links in bipartite networks of papers and their cited sources or in tripartite networks of papers, cited sources, and terms used in papers and sources. HLC only assumes that a source is cited for only one reason or for only very few similar reasons in one paper. In the case of terms, the assumption is that authors use one term in one paper with only one meaning. These assumptions are both not only very plausible but could even be tested in case studies. A further advantage of link clustering is that it allows to combine citation and textual information in tripartite graphs---a very `natural' solution of this longstanding problem (cf. e.g. the introduction of reference \cite{janssens2009hybrid} and sources cited there).

For \textbf{MONC} there are no further assumptions beyond the four mentioned above and the one about paper-similarity networks. However, we found that MONC needs some post-processing to reveal the hierarchy of a graph. Thus, it is assumed that hierarchical clustering of the nodes' perspectives results in a realistic hierarchy of topics.

\subsection{Methodological Aspects}

Our implementations of the three approaches to overlapping communities all involve a hard clustering procedure. The fuzzification algorithm uses hard clusters of nodes as input, i.e. clustering has to be done as pre-processing. Hard clustering of fuzzy natural communities is part of MONC's post-processing. In the case of HLC, the agorithm itself is a hard clustering procedure. For HLC we only need to calculate link similarities as some kind of pre-processing.

We have presented results obtained with only one standard hard-cluster algorithm per approach but tested also other ones. Fuzzification and link clustering also worked with Louvain algorithm \cite{blondel2008fast} but we abandoned this fast modularity-driven method due to its use of global information (and the poor hierarchical structure obtained). Fuzzification could perform better with average-linkage clustering but its dendrogram showed only very small stable communities. In the case of average link clustering (HLC) we succeeded in finding relatively stable communities of some size after re-parametrising the dendrogram's similarity axis.

When it comes to defining grades of a node's memberships in different communities, link clustering implies a very plausible and consistent definition. MONC membership grades could also be defined alternatively to the ansatz used here (equation \ref{def.mu}). We see this methodological ambiguity as a disadvantage (arbitrary parameters are only a special case of such an ambiguity). In our fuzzification experiments we calculated fractional membership grades using non-fractional (zero or full) membership of nodes as input (equation \ref{def.fuzzy.mu}). This inconsistent definition could possibly be avoided  by an iterative algorithm that in each step uses the fractional grades as input to calculate new ones. We did not yet test such an iteration procedure and hence do not know whether it would converge or not. An alternative would be to use the fitness balances of a node as input for a membership definition. Our attempts to define grades this way led to communities with only very few full members, which could be a desired feature for topic extraction that cannot be achieved by HLC membership grades.

MONC membership grades fit into the framework of fuzzy set theory because a node's grades in general do not sum up to unity. Link clustering leads to node grades which are normalised. Thus, an HLC grade is more adequately interpreted as a probability.

\section{Discussion}

We implemented three local approaches to the identification of overlapping and hierarchically ordered communities in networks as algorithms and tested their ability to extract manually defined thematic substructures from a network of information-science papers and their cited sources.

Hierarchical clustering of citation links proved to be the most satisfactory approach---with regard to the test results, to its methodological simplicity, to its ability to work with the broadest information basis (the bipartite graph of papers and sources), and to its potential for a simple inclusion of text information in addition to citation data---an issue on top of our agenda.

Clustering citation links does not need to be restricted to a small period of time but can also be applied to a longer time period. This might make it possible to solve the problem of tracing the development of topics over time. The only limitation HLC encounters is the limited coverage of publication databases, i.e. the existence 
of citation links to publications that are not included in the database.

MONC was found to be useful for overcoming the longstanding problem of field delineation by greedily expanding the paper set downloaded from a citation database \cite[p. 19]{havemann2011identification}. Instead of delineating research fields by journal sets, they can be identified with a large enough natural community---obtained with low enough resolution---of an appropriate seed node. 

Hierarchical clustering of citation links can be applied to this problem too. We only have to cluster the environment of a seed set and then to omit all branches in the dendrogram which are not sub-branches of the most stable community. The next iteration starts from this reduced paper set. As in the MONC case, we would alternate between expansion (inclusion of neighbours) and partial reduction of the sample (exclusion of neighbours which do not improve community fitness or---in the HLC case---stability).

While the fuzzification algorithm only sometimes creates good clusters in terms of target topics, iterating fitness-based optimisation may lead to more consistent clusters by removing loosely connected nodes. If the iteration is done node by node
this leads to a version of the LFM algorithm \cite{lancichinetti2009detecting} applied to hard clusters instead of single nodes or cliques, an approach already proposed by Baumes \textit{et al.} (2005) \cite{baumes2005efficient}.

\section*{Acknowledgements}
This work is part of a project in which we develop methods for measuring the diversity of research. The project is funded by the German Ministry for Education and Research (BMBF). We would like to thank all developers of \textbf{R}.\footnote{\url{http://www.r-project.org}}

\section*{Author Contributions}
Conceived and designed the experiments: all authors. Performed the experiments: AS (fuzzification), MH (link clustering), FH (MONC). Analysed the data: AS (fuzzification), MH (link clustering, comparison), FH (MONC). Wrote the paper: FH, JG. Discussed the text: all authors. 

 % The bibtex filename
\bibliography{informetrics}

\begin{thebibliography}{10}
\providecommand{\url}[1]{\texttt{#1}}
\providecommand{\urlprefix}{URL }
\expandafter\ifx\csname urlstyle\endcsname\relax
  \providecommand{\doi}[1]{doi:\discretionary{}{}{}#1}\else
  \providecommand{\doi}{doi:\discretionary{}{}{}\begingroup
  \urlstyle{rm}\Url}\fi
\providecommand{\bibAnnoteFile}[1]{%
  \IfFileExists{#1}{\begin{quotation}\noindent\textsc{Key:} #1\\
  \textsc{Annotation:}\ \input{#1}\end{quotation}}{}}
\providecommand{\bibAnnote}[2]{%
  \begin{quotation}\noindent\textsc{Key:} #1\\
  \textsc{Annotation:}\ #2\end{quotation}}
\providecommand{\eprint}[2][]{\url{#2}}

\bibitem{vanRaan2004measuring}
Van~Raan A (2004) Measuring science.
\newblock In: Moed HF, Gl{\"a}nzel W, Schmoch U, editors, Handbook of
  quantitative science and technology research: The use of publication and
  patent statistics in studies of S\&T systems, Dordrecht etc.: Kluwer,
  chapter~11. pp. 19--50.
\bibAnnoteFile{vanRaan2004measuring}

\bibitem{Zitt2005relativity}
Zitt M, Ramanana-Rahary S, Bassecoulard E (2005) Relativity of citation
  performance and excellence measures: {From} cross-field to cross-scale
  effects of field-normalisation.
\newblock Scientometrics 63: 373--401.
\bibAnnoteFile{Zitt2005relativity}

\bibitem{janssens2008hmi}
Janssens F, Gl{\"a}nzel W, {De Moor} B (2008) {A hybrid mapping of information
  science}.
\newblock Scientometrics 75: 607--631.
\bibAnnoteFile{janssens2008hmi}

\bibitem{klavans2011using}
Klavans R, Boyack K (2011) Using global mapping to create more accurate
  document-level maps of research fields.
\newblock Journal of the American Society for Information Science and
  Technology 62: 1--18.
\bibAnnoteFile{klavans2011using}

\bibitem{sullivan1977cocitation}
Sullivan D, White D, Barboni E (1977) Co-citation analyses of science: An
  evaluation.
\newblock Social Studies of Science 7: 223--240.
\bibAnnoteFile{sullivan1977cocitation}

\bibitem{amsterdamska1989citations}
Amsterdamska O, Leydesdorff L (1989) {Citations: Indicators of significance?}
\newblock Scientometrics 15: 449--471.
\bibAnnoteFile{amsterdamska1989citations}

\bibitem{marshakova1973ssm}
Marshakova IV (1973) {Sistema svyazey mezhdu dokumentami, postroyennaya na
  osnove ssylok (po ukazatelyu ``Science Citation Index``)}.
\newblock Nauchno-Tekhnicheskaya Informatsiya Seriya 2 -- Informatsionnye
  Protsessy i Sistemy 6: 3--8.
\bibAnnoteFile{marshakova1973ssm}

\bibitem{Small1973cocitation}
Small H (1973) {Co-citation in the Scientific Literature: A New Measure of the
  Rela\-tionship Between Two Documents}.
\newblock Journal of the American Society for Information Science 24: 265--269.
\bibAnnoteFile{Small1973cocitation}

\bibitem{Mitesser2008mdr}
Mitesser O, Heinz M, Havemann F, Gl{\"a}\-ser J (2008) {Measuring Diversity of
  Research by Extracting Latent Themes from Bipartite Networks of Papers and
  References}.
\newblock In: Kretschmer H, Havemann F, editors, Proceedings of WIS 2008:
  Fourth International Conference on Webometrics, Informetrics and
  Scientometrics \& Ninth COLLNET Meeting. Humboldt-Universit{\"a}t zu Berlin,
  Berlin: Gesellschaft f{\"u}r Wissenschaftsforschung.
\newblock ISBN: 978-3-934682-45-0, \url{http://www. collnet.
  de/Berlin-2008/MitesserWIS2008mdr. pdf}.
\bibAnnoteFile{Mitesser2008mdr}

\bibitem{Lancichinetti2011finding}
Lancichinetti A, Radicchi F, Ramasco JJ, Fortunato S (2011) Finding
  statistically significant communities in networks.
\newblock PLoS ONE 6: e18961.
\bibAnnoteFile{Lancichinetti2011finding}

\bibitem{glaser2006wissenschaftliche}
Gl{\"a}ser J (2006) {Wissenschaftliche Produktionsgemeinschaften: Die soziale
  Ordnung der Forschung}.
\newblock Campus Verlag.
\bibAnnoteFile{glaser2006wissenschaftliche}

\bibitem{fortunato2010community}
Fortunato S (2010) Community detection in graphs.
\newblock Physics Reports 486: 75--174.
\bibAnnoteFile{fortunato2010community}

\bibitem{wang2009adjusting}
Wang X, Jiao L, Wu J (2009) {Adjusting from disjoint to overlapping community
  detection of complex networks}.
\newblock Physica A: Statistical Mechanics and its Applications 388:
  5045--5056.
\bibAnnoteFile{wang2009adjusting}

\bibitem{ahn2010link}
Ahn Y, Bagrow J, Lehmann S (2010) Link communities reveal multiscale complexity
  in networks.
\newblock Nature 466: 761--764.
\bibAnnoteFile{ahn2010link}

\bibitem{lancichinetti2009detecting}
Lancichinetti A, Fortunato S, Kertesz J (2009) {Detecting the overlapping and
  hierarchical community structure in complex networks}.
\newblock New Journal of Physics 11: 033015.
\bibAnnoteFile{lancichinetti2009detecting}

\bibitem{zachary1977information}
Zachary W (1977) An information flow model for conflict and fission in small
  groups.
\newblock Journal of Anthropological Research 33: 452--473.
\bibAnnoteFile{zachary1977information}

\bibitem{Tibely2011Criterions}
Tib{\'e}ly G (2011) Criterions for locally dense subgraphs.
\newblock Arxiv preprint arXiv:11033397 .
\bibAnnoteFile{Tibely2011Criterions}

\bibitem{friggeri2011ego}
Friggeri A, Chelius G, Fleury E (2011) Ego-munities, exploring socially
  cohesive person-based communities.
\newblock Arxiv preprint arXiv:11022623 .
\bibAnnoteFile{friggeri2011ego}

\bibitem{Radicchi2004defining}
Radicchi F, Castellano C, Cecconi F, Loreto V, Parisi D (2004) Defining and
  identifying communities in networks.
\newblock Proceedings of the National Academy of Sciences of the United States
  of America 101: 2658--2663.
\bibAnnoteFile{Radicchi2004defining}

\bibitem{gregory2011fuzzy}
Gregory S (2011) Fuzzy overlapping communities in networks.
\newblock Journal of Statistical Mechanics: Theory and Experiment 2011: P02017.
\bibAnnoteFile{gregory2011fuzzy}

\bibitem{baumes2005finding}
Baumes J, Goldberg M, Krishnamoorthy M, Magdon-Ismail M, Preston N (2005)
  {Finding communities by clustering a graph into overlapping subgraphs}.
\newblock In: International Conference on Applied Computing (IADIS 2005).
\bibAnnoteFile{baumes2005finding}

\bibitem{baumes2005efficient}
Baumes J, Goldberg M, Magdon-Ismail M (2005) {Efficient identification of
  overlapping communities}.
\newblock Intelligence and Security Informatics 3495: 27--36.
\bibAnnoteFile{baumes2005efficient}

\bibitem{clauset2005flc}
Clauset A (2005) {Finding local community structure in networks}.
\newblock Physical Review E 72: 26132.
\bibAnnoteFile{clauset2005flc}

\bibitem{lee2011seeding}
Lee C, Reid F, McDaid A, Hurley N (2011) Seeding for pervasively overlapping
  communities.
\newblock Arxiv preprint arXiv:11045578 .
\bibAnnoteFile{lee2011seeding}

\bibitem{evans2009line}
Evans T, Lambiotte R (2009) {Line graphs, link partitions, and overlapping
  communities}.
\newblock Physical Review E 80: 16105.
\bibAnnoteFile{evans2009line}

\bibitem{Ball2011efficient}
{Ball} B, {Karrer} B, {Newman} M (2011) {An efficient and principled method for
  detecting communities in networks.}
\newblock ArXiv preprint arXiv11043590 .
\bibAnnoteFile{Ball2011efficient}

\bibitem{kim2011map}
Kim Y, Jeong H (2011) The map equation for link community.
\newblock Arxiv preprint arXiv:11050257 .
\bibAnnoteFile{kim2011map}

\bibitem{rosvall2008maps}
Rosvall M, Bergstrom C (2008) Maps of random walks on complex networks reveal
  community structure.
\newblock Proceedings of the National Academy of Sciences 105: 1118.
\bibAnnoteFile{rosvall2008maps}

\bibitem{Ghosh2011Identifying}
Ghosh S, Kane P, Ganguly N (2011) {Identifying overlapping communities in
  folksonomies or tripartite hypergraphs, WWW 2011, Poster, March 28–April 1,
  2011, Hyderabad, India} : 39--40.
\bibAnnoteFile{Ghosh2011Identifying}

\bibitem{havemann2011identification}
Havemann F, Heinz M, Struck A, Gl{\"a}ser J (2011) {Identification of
  Overlapping Communities by Locally Calculating Community-Changing Resolution
  Levels}.
\newblock Journal of Statistical Mechanics: Theory and Experiment 2011: P01023.
\bibAnnoteFile{havemann2011identification}

\bibitem{lee2010detecting}
Lee C, Reid F, McDaid A, Hurley N (2010) {Detecting highly overlapping
  community structure by greedy clique expansion}.
\newblock In: Proceedings of the 4th SNA-KDD Workshop.
\newblock ArXiv: \url{http://arxiv.org/abs/1002.1827}.
\bibAnnoteFile{lee2010detecting}

\bibitem{janssens2009hybrid}
Janssens F, Zhang L, Moor B, Gl{\"a}nzel W (2009) Hybrid clustering for
  validation and improvement of subject-classification schemes.
\newblock Information Processing \& Management 45: 683--702.
\bibAnnoteFile{janssens2009hybrid}

\bibitem{blondel2008fast}
Blondel V, Guillaume J, Lambiotte R, Lefebvre E (2008) Fast unfolding of
  communities in large networks.
\newblock Journal of Statistical Mechanics: Theory and Experiment 2008: P10008.
\bibAnnoteFile{blondel2008fast}

\end{thebibliography}

\end{document}